\documentclass{aa}  
\usepackage{graphicx}
\usepackage{txfonts}

\usepackage[usenames,dvipsnames,svgnames,table]{xcolor}
\usepackage{float}
\usepackage[
breaklinks=true,
hyperindex=true,
colorlinks=true,
linkcolor=Sepia,
citecolor=Purple]{hyperref}
\usepackage{bm}
\usepackage{multirow}
\usepackage{booktabs}
\usepackage{arydshln}

\bibpunct{(}{)}{;}{a}{}{,} 

\definecolor{ogreen}{rgb}{0.07,0.54,0.03}

\newcommand{\E}[1]{\times 10^{#1}}
 
\newcommand{\kms}{~\mathrm{km}~\mathrm{s}^{-1}}
\newcommand{\kpc}{~\mathrm{\mathrm{kpc}}}

\makeatletter
\renewcommand*\aa@pageof{, page \thepage{} of \pageref*{LastPage}}
\makeatother

\titlerunning{The inefficient high Mach regime for star formation. II: Numerical simulations}
\begin{document} 

\graphicspath{{./figures/}{}}

\defcitealias{hennebelleInefficientStarFormation2024}{Paper~I}

\title{Inefficient star formation in high Mach number environments}
\subtitle{II. Numerical simulations and comparison with analytical models}

   \author{Noé Brucy \inst{\ref{inst1}} 
   \and  Patrick Hennebelle\inst{\ref{inst2}} 
   \and  Tine Colman\inst{\ref{inst2}} 
   \and Ralf S.\ Klessen\inst{\ref{inst1},\ref{inst4}} 
    \and Corentin Le Yhuelic\inst{\ref{inst2},\ref{inst3}} 
    }

   \institute{ Universität Heidelberg, Zentrum für Astronomie, Institut für Theoretische Astrophysik, Albert-Ueberle-Str 2, D-69120 Heidelberg, Germany,
      \label{inst1} 
      \and 
     Université Paris-Saclay, Université Paris Cité, CEA, CNRS, AIM, 91191, Gif-sur-Yvette, France,
      \label{inst2}
      \and 
      Université Paris-Saclay, ENS Paris Saclay, 91191, Gif-sur-Yvette, France,
      \label{inst3} 
      \and
      Universit\"{a}t Heidelberg, Interdisziplin\"{a}res Zentrum f\"{u}r Wissenschaftliches Rechnen, Im Neuenheimer Feld 205, 69120 Heidelberg, Germany
      \label{inst4}
}

\date{Received \today; submitted to A\&A}

  \abstract
{Predicting the star formation rate (SFR) in galaxies is crucial to understand their evolution and morphology.
To do so requires a fine understanding of how dense structures of gas are created and collapse. 
In that, turbulence and gravity play a major role.}
{Within the gravo-turbulent framework, we assume that turbulence shapes the interstellar medium, creating density fluctuations that, if gravitationally unstable, 
will collapse and form stars. 
The goal of this work is to quantify how different regimes of turbulence,
characterized by the strength and compressibility of the driving,
shape the density field. 
We are interested in the outcome in terms of SFR and how it compares with existing analytical models for the SFR.}
{We run a series of hydrodynamical simulations of turbulent gas. The simulations are first conducted without gravity, 
so that the density and velocity are shaped by the turbulence driving. Gravity is then switched on, and the SFR is measured 
and compared with analytical models. The physics included in these simulations is very close to the one assumed in the classical 
gravo-turbulent SFR analytical models, which makes the comparison straightforward.}
{We found that the existing analytical models convincingly agree with simulations at low Mach number, but we measure a much lower SFR 
in the simulation with a high Mach number. We develop, in a companion paper, an updated physically motivated SFR 
model that reproduces well the inefficient high Mach regime of the simulations.}
{Our work demonstrates that accurate estimations of the turbulent-driven replenishment time of dense structures and the dense gas spatial distribution 
are necessary to correctly predict the SFR in the high Mach regime.
The inefficient high-Mach regime is a possible explanation for the low SFR found in dense and turbulent environments such as the centre of our Milky Way and other galaxies.}


   \maketitle

%
\section{Introduction}
\label{sec:intro}

How fast a self-gravitating gaseous system converts its mass into stars is a
major question that needs to be addressed to understand the history of our
Universe. Determining the star formation rate (SFR), which is the mass of
stars formed per unit of time, has led to a vast number of observational studies
\citep[e.g.][]{schmidt1959,lada2010,kennicutt2012,schrubaHowGalacticEnvironment2019,sunStarFormationLaws2023a}.
From the theoretical side, numerous
efforts have also been dedicated to this question, either using analytical
models or performing numerical simulations.

The analytical models, more extensively described in our companion paper \citep{hennebelleInefficientStarFormation2024}, henceforth referred to as \citetalias{hennebelleInefficientStarFormation2024}, are all based on
an estimate of the fraction of self-gravitating gas and a characteristic time,
usually the freefall time, in which the dense gas eventually collapses. For
instance, several models consider a log-normal probability density function (PDF), sometimes with a power-law tail, and integrate the fraction
of gas that is above a given density threshold
\citep{krum2005,padoanStarFormationRate2011,hennebelleAnalyticalStarFormation2011,renaud2012, padoanSimpleLawStar2012,burkhartStarFormationRate2018}.
This threshold is usually determined by invoking a density or a scale for gravitational
instability.

Turbulence is thought to be a key player in the generation of self-gravitating structures.
From the numerical simulation side, studies that investigate the role of turbulence on the SFR can be divided into two types: simulations with a prescribed large-scale driving and simulations that explicitly treat various stellar feedback processes.
When large-scale driving is employed
\citep{maclowControlStarFormation2004,padoanStarFormationRate2011,
federrathStarFormationRate2012,federrathStarFormationEfficiency2013}, the goal
has usually been to understand basic physical principles in a relatively
idealised setup.
Typical approximations in the numerical models are the consideration of isothermal gas and periodic boundary conditions.
Detailed comparisons between the various
analytical predictions and the simulation results have been
performed by \citet{padoanStarFormationRate2011} and
\citet{federrathStarFormationRate2012}, concluding that, by adjusting some
fudge factors, good agreement between models and simulations has been inferred.

Simulations that include stellar feedback can be further divided into two
categories. The first category of simulations considers relatively large
computational boxes of several hundreds of parsecs and mainly relies on
supernova remnants to limit star formation
\citep[e.g.][]{kimRegulationStarFormation2011,
hennebelleSimulationsMagnetizedMultiphase2014,
walchSILCCSImulatingLifeCycle2015,padoanSupernovaDrivingIII2016,
iffrigStructureDistributionTurbulence2017,
kimThreephaseInterstellarMedium2017}. These simulations aim to represent a
galaxy portion containing an interstellar medium (ISM) in which stellar feedback self-regulates star formation.
These studies, which have mainly considered column densities typical of
the Milky Way, have generally concluded that supernova explosions, sometimes
coupled with HII regions \citep{collingImpactGalacticShear2018} and/or stellar
winds \citep{gattoSILCCProjectIII2017}, reduce the SFR by one to two orders of
magnitude and are compatible with the ones inferred from observations. The
second category of simulations focuses on smaller computational boxes, with a size less than a hundred parsecs, and consider the effect of protostellar jets and HII regions
\citep{Wang10,federrathDensityStructureStar2015,verliat2022}. Here as well,
stellar feedback appears to be able to substantially reduce the SFR which without feedback would be too high compared to observations.

Recently, \citet{brucyLargescaleTurbulentDriving2020, brucyLargescaleTurbulentDriving2023} and 
\citet{colmanSignatureLargescaleTurbulence2022} have carried out a series of
simulations in which both large-scale turbulent driving and stellar feedback
are considered. Apart from the fact that large-scale turbulence is indeed
expected from large-scale gravitational instability
\citep[e.g.][]{wadaGravitydrivenTurbulenceGalactic2002},
\citet{brucyLargescaleTurbulentDriving2020} have concluded that the
relation between gas column density and SFR, known as the 
Schmidt-Kennicutt relation \citep{kennicutt2012}, cannot be reproduced by
simulations that consider only stellar feedback, in particular because at high
column density the inferred star formation rate is far too high. 
This is a conclusion
that seems to be compatible with the results presented by
\citet{kimFirstResultsSMAUG2020}, but possibly in tension with what is
reported by \citet{rathjenSILCCVIIGas2023}, who find lower values of the SFR than the two aforementioned studies when stronger magnetic fields are considered.
\citet{brucyLargescaleTurbulentDriving2020} found that it is possible to
reproduce the Schmidt-Kennicutt relation when the turbulent driving is intense
enough. \citet{brucyLargescaleTurbulentDriving2023} looked in more details at the effects of the strength and compressibility of the turbulent driving on the SFR and showed that a fairly solenoidal driving associated with a velocity dispersion of 30$ \kms$ can already reduce the SFR by one order of magnitude.

Numerical results thus present clear evidence that the SFR is 
regulated by a form of injection of 
kinetic energy into the star-forming systems. The exact mechanism(s) through which this operates is debated.
Whether this kinetic energy takes
the form of, or is equivalent to, a turbulent cascade remains to be clarified.
At high column densities, the velocity dispersion 
required to obtain SFRs compatible with observed values can be as high
as 100$ \kms$.
Such values may indeed have been observed in high column
density galaxies \citep[e.g.][]{swinbankPropertiesStarformingInterstellar2012}.

Assuming a sound speed of 0.3$ \kms$, such high velocity dispersions correspond to Mach numbers exceeding 100.
The existing SFR theories have never been compared to simulations with such
high Mach numbers. Moreover, as discussed in \citetalias{hennebelleInefficientStarFormation2024}, the existing
models present debatable physical assumptions, suggesting room for improvement. The main shortcomings of the previous models are:
\begin{enumerate}[i.]
    \item they rely on the density PDF only and do not account for the spatial
    distribution of the density field,
    \item they generally integrate the density PDF over a density threshold
    that is related to a single Jeans or sonic length and do not account for
    the scale-dependent turbulent velocity dispersion,
    \item they rely on the freefall time as the characteristic time scale
    over which the unstable gas disappears even though the time needed to
    replenish the dense gas after it has collapsed may actually be longer.
\end{enumerate}
In the present paper, we present a large suite of 
turbulent-driven
numerical simulations in which
we systematically explore 
a large range of Mach numbers, including 
the
high Mach number regime. We confront the results both with several existing
models as well as with the new \emph{Turbulent Support} (TS) model developed in \citetalias{hennebelleInefficientStarFormation2024}, finding
remarkable agreement with the latter. Importantly, we identify a new regime for
high Mach numbers in which star formation is very inefficient.

This paper is structured as follows.
In the second section, we present our numerical setup. In the third section, a large suite of numerical simulations
without self-gravity is presented. The goal is to measure the flow statistical properties that must be known to 
apply the analytical model. In the fourth section, we present the same suite of simulations but this time self-gravity 
and sink particles are employed and the SFR is inferred. 
We then compare the numerical results with 
the analytical models.
In the fifth section, we take a step back and analyses the caveats of our study. 
The sixth section concludes the paper. 
The main notations used in the article are presented in Table \ref{tbl:notation}.

\begin{table}[ht]
	\caption{Main notations used in the article.}
  \begin{center}
	\begin{tabular}{lp{6cm}}
	\toprule
	Notation &  Description \\
	\midrule
	$G$                       & Gravitational constant \\
    $m_p$                     & Mass of a proton \\
    $\mu = 1.4$               & Mean molecular weight, in unit of $m_p$ \\
    \midrule
    $L_\mathrm{box}$          & Size of the simulation box \\
    $N_x$                     & Number of cells on one side  of the cubic domain \\ 
    \midrule
    $L_i$                     & Injection scale of the turbulence \\
    $f_\mathrm{rms}$          & Strength of the driving \\
    $\chi$          & Compressive fraction of the driving \\
    $T_\mathrm{driv}$         & Autocorrelation time of the turbulence\\
    \midrule
    $\sigma$                  & 3D mass-weighted velocity dispersion\\
    $\mathcal{M}$             & 3D Mach number\\
    \midrule
    $\rho_0$ & Initial density  \\
    $\delta = \ln (\rho / \rho_0) $ & Natural logarithm of the normalized density \\
    $S_{\delta}$             & Volume weighted variance of the ln-density $\delta$ \\
    $T_{\mathrm{CH}}$         & $T$ parameter of \cite{hopkinsModelNonlognormalDensity2013} \\
    $n_d$              & Slope of the ln-density power spectrum \\
    $\eta_d$ & $\eta_d = (n_d-3) / 2$ \\

	\bottomrule
	\end{tabular} 
  \end{center}
	\label{tbl:notation}
\end{table}

\begin{table*}[ht]
\caption{Simulations}\label{tbl:turbox_simu}
\begin{tabular}{clccccccccc}
\toprule
Group & Name &    $L_\mathrm{box}$ &   $N_x$  &   $\chi$ &    $f_\mathrm{rms}$ &   $\mathcal{M}$  &  $S_{\delta}$ &     $\eta_d$ \\
 &  &    [pc] &   &  & {\scriptsize $[1.46 \E{-4}~\mathrm{km}\cdot\mathrm{s}^{-1}\cdot\mathrm{Myr}^{-1} ]$}  &   &  &  &  \\
\midrule
\multirow{ 8}{*}{L200\_N512}  
  & L200\_N512\_comp0.5\_frms50 &  200  & 512 &  0.5 &      50 &    2.85 &        1.59 &     0.38 \\
  & L200\_N512\_comp0.5\_frms1e2 &  200  & 512 &  0.5 &     100 &   4.02 &        2.29 &      0.4 \\
  & L200\_N512\_comp0.5\_frms5e2 &  200  & 512 &  0.5 &     500 &    8.97 &        4.31 &     0.34 \\
  & L200\_N512\_comp0.5\_frms1e3 &  200  & 512 &  0.5 &    1000 &    12.6 &        5.44 &    0.232 \\
  & L200\_N512\_comp0.5\_frms5e3 &  200  & 512 &  0.5 &    5000 &    28.4 &         9.1 &   0.0783 \\
  & L200\_N512\_comp0.5\_frms1e4 &  200  & 512 &  0.5 &   10000 &   40.1 &          10 & -0.00884 \\
  & L200\_N512\_comp0.5\_frms5e4 &  200  & 512 &  0.5 &   50000 &     89.3 &        11.4 &  -0.0406 \\
\midrule
\multirow{3}{*}{L1000\_N256}
  & L1000\_N256\_comp0.5\_frms1e2&  1000 &  256 &  0.5 &    100  & 9.3 &        4.22 &  0.352 \\
  & L1000\_N256\_comp0.5\_frms1e3 &  1000 &  256 &  0.5 &   1000  &  29 &        8.73 &  0.178 \\
  & L1000\_N256\_comp0.5\_frms1e4 &  1000 &  256 &  0.5 &  10000  &   91.5 &        10.9 &  0.134 \\
\midrule

\multirow{16}{*}{L1000\_N512}
  & L1000\_N512\_comp0\_frms1e2  & 1000 &  512 &  0.0 &    100 &    8.7 &         3.7 &    0.248 \\
  & L1000\_N512\_comp0\_frms1e3  & 1000 &  512 &  0.0 &    100 &    27.6 &        7.89 &  -0.0179 \\
  & L1000\_N512\_comp0\_frms5e3  & 1000 &  512 &  0.0 &   1000 &    61.5 &        10.2 &   -0.055 \\
  & L1000\_N512\_comp0\_frms1e4  & 1000 &  512 &  0.0 &  10000 & 87.3 &        10.4 &  -0.0749 \\
  & L1000\_N512\_comp0\_frms5e4  & 1000 &  512 &  0.0 &  50000 &   123 &        10.8 &   -0.046 \\
\cmidrule[0.01em](r){2-10}
  & L1000\_N512\_comp0.5\_frms50  &  1000 &  512 &  0.5 &     50 &    6.36 &        3.25 &    0.322 \\
  & L1000\_N512\_comp0.5\_frms1e2 &  1000 &  512 &  0.5 &    100 & 8.88 &        4.46 &    0.329 \\
  & L1000\_N512\_comp0.5\_frms2e2 &  1000 &  512 &  0.5 &    200 &  12.6 &        5.62 &     0.27 \\
  & L1000\_N512\_comp0.5\_frms5e2 &  1000 &  512 &  0.5 &    500 &   20 &        7.83 &    0.132 \\
  & L1000\_N512\_comp0.5\_frms1e3 &  1000 &  512 &  0.5 &   1000 &   28.4 &         9.2 &   0.0497 \\
  & L1000\_N512\_comp0.5\_frms5e3 &  1000 &  512 &  0.5 &   5000 &  263.6 &        11.7 &   0.0124 \\
  & L1000\_N512\_comp0.5\_frms1e4 &  1000 &  512 &  0.5 &  10000 &    90 &          12 &  -0.0399 \\
  & L1000\_N512\_comp0.5\_frms5e4 &  1000 &  512 &  0.5 &  50000 &   127 &          12 &  -0.0298 \\
\cmidrule[0.01em](r){2-10}
  & L1000\_N512\_comp1\_frms1e2  & 1000 &  512 &  1.0 &    100 &    8.31 &        5.28 &    0.401 \\ 
  & L1000\_N512\_comp1\_frms1e3  & 1000 &  512 &  1.0 &   1000 &   26.9 &        10.8 &    0.165 \\
  & L1000\_N512\_comp1\_frms1e4  & 1000 &  512 &  1.0 &  10000 &  84.9 &        13.6 &   0.0436 \\
  & L1000\_N512\_comp1\_frms5e4  & 1000 &  512 &  1.0 &  50000 &   191 &        13.9 &   0.0402 \\
\midrule
\multirow{3}{*}{L1000\_N1024}
  & L1000\_N1024\_comp0.5\_frms1e2&  1000 &  1024 &  0.5 &    100  &  9.01 &        4.28 &   0.329 \\
  & L1000\_N1024\_comp0.5\_frms1e3 &  1000 &  1024 &  0.5 &   1000  &  28.8 &        9.47 &  0.0505 \\
  & L1000\_N1024\_comp0.5\_frms1e4 &  1000 &  1024 &  0.5 &  10000  &  91.4 &        12.7 & -0.0739 \\
\bottomrule
\end{tabular}

\tablefoot{ For each simulation we display its group, its name, the size of the simulation domain $L_\mathrm{box}$ in pc, the resolution, expressed as the number of cells $N_x$ on one side of the cubic simulation box, the driving compressibility fraction $\chi$, the driving strength $f_\mathrm{rms}$  in code units corresponding to an RMS acceleration of $1.46 \ \kms\cdot\mathrm{Myr}^{-1}$, the measured Mach number $\mathcal{M}$, the measured dispersion of the natural logarithm of density $S_{\delta}$ and the value of the $\eta_d$ parameter as computed from the slope of the ln-density power spectrum. All simulations have an initial number density of $n_0 = 1.5~\mathrm{cm^{-3}}$.}

\end{table*}

\begin{figure}
    \centering
    \includegraphics[width=\linewidth]{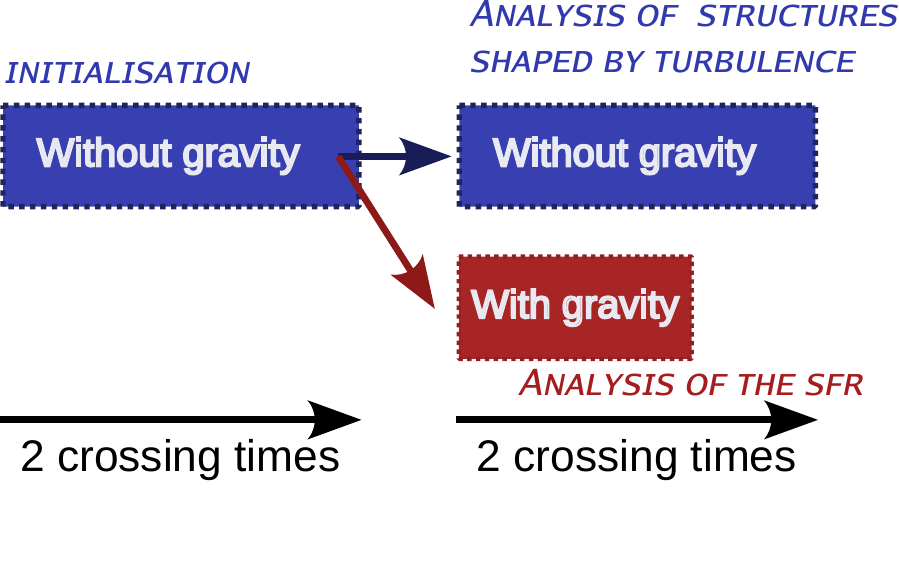}
    \caption{Illustration of the simulation procedure.}
    \label{fig:process}
\end{figure}

\begin{figure*}
    \centering
    \includegraphics[width=0.8\textwidth]{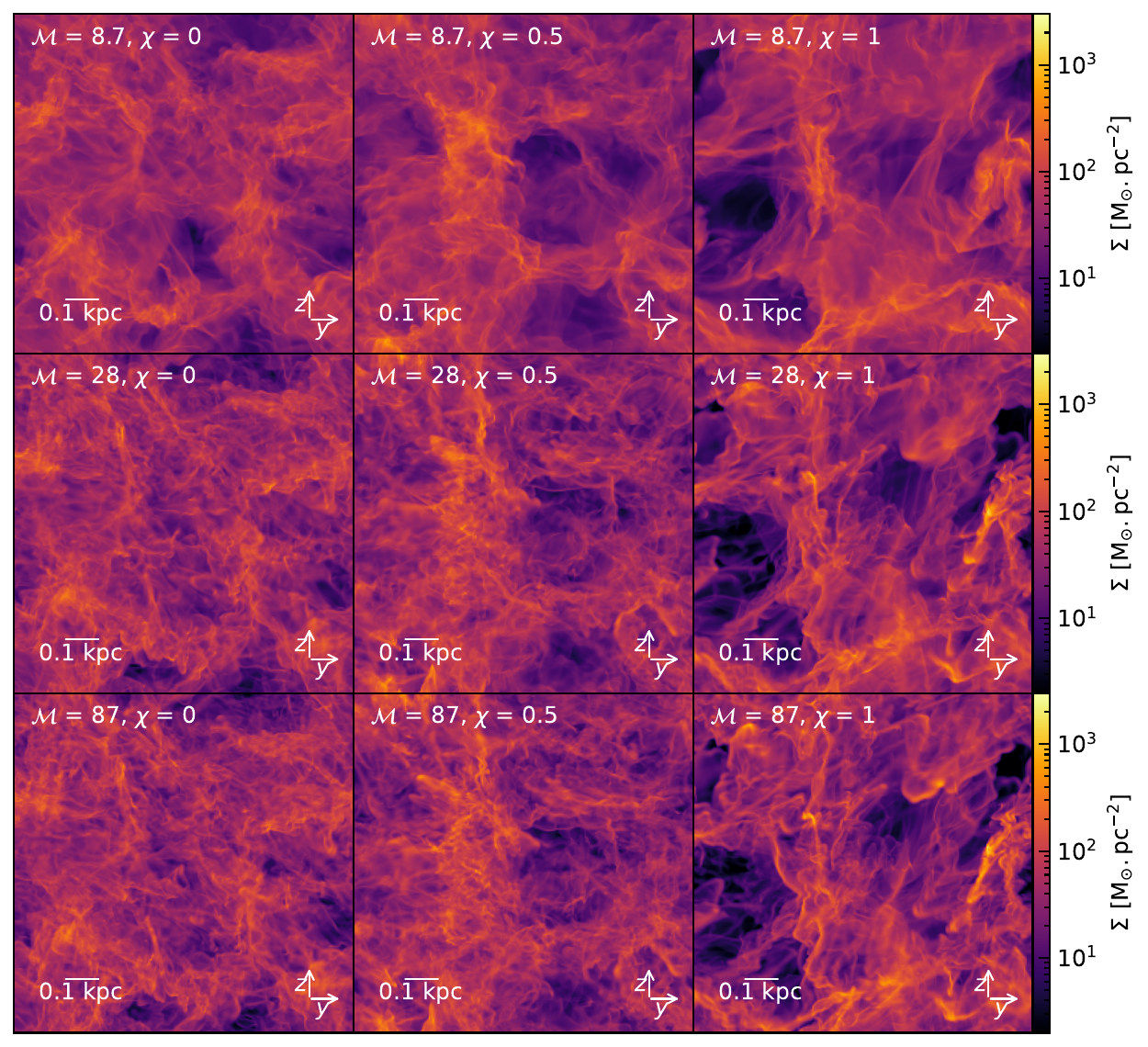}
    \caption{Column density for the simulations without gravity (group L1000\_N512), after two autocorrelation time of the turbulent driving $t = 2 T_\mathrm{driv}$. 
    Compressibility increases from left ($\chi = 0$, purely solenoidal) to right ($\chi = 1$, purely compressive). Driving strength increases from the top to the bottom.}
    \label{fig:coldens_nograv}
\end{figure*}

\section{Numerical setup}
\label{sec:setup}

\subsection{Code and numerical parameters}
We run hydrodynamical simulations of turbulent gas with the \textsc{ramses} code 
\citep{teyssierCosmologicalHydrodynamicsAdaptive2002}. A Godunov scheme is used, 
with the HLLC Riemann solver and the MinMod slope limiter. A 
uniform computational grid is employed to avoid any issues with adaptive mesh 
refinement, and the spatial resolution is varied from 256$^3$ to 512$^3$ and up to 1024$^3$ 
for a few runs.

\subsection{Initial conditions}
The initial conditions are chosen to be as simple as possible. They essentially 
consist of a uniform density field subject to large-scale turbulent driving. The 
simulation domain is a periodic cubic box with a size length $L_\mathrm{box}$ of
$1 \kpc$. It is initially filled with a uniform distribution 
of gas with a density  $\rho_0 =  1.5\cdot \mu\ m_p\ \mathrm{cm}^{-3}$. The chemical 
evolution of the gas is not tracked, and the gas is assumed to have a mean 
molecular weight $\mu =1.4$ in atomic mass unit $m_p$. The gas is isothermal with 
a temperature of 10 K. These parameters are motivated by recent studies 
\citep{brucyLargescaleTurbulentDriving2020,brucyLargescaleTurbulentDriving2023} 
where a drastic decrease in SFR has been observed, depending on initial densities, 
for velocity dispersions on the order of several tens to one hundred~km~s$^{-1}$.
These isothermal models have no characteristic scale. As a consequence, the significant 
parameters are the Mach number and the box size to Jeans length ratio, 
or equivalently, the number of Jeans masses in the computational box.

Our simulation process is illustrated in Fig.~\ref{fig:process}.
Each simulation is run without gravity for two crossing times in order to  fully developed the turbulence.
It is then further run for two additional crossing times without gravity, where the properties of the generated density and velocity fields are studied with enough statistics (see Section~\ref{sec:nograv}). 
It is also run in parallel with gravity, to compute the resulting SFR (see Section~\ref{sec:grav}). 
In all simulations, turbulence 
is driven as described in Section \ref{subsec:turbinj}. 

This procedure 
aligns with the goal and essence of analytical models, predicting the SFR based 
on parameters like mean density, temperatures, Mach numbers, and turbulent 
injection scale. The amount of self-gravitating gas is a prediction of these models.
The parameters used for each simulation are listed in Table~\ref{tbl:turbox_simu}.

\subsection{Injection of turbulence}
\label{subsec:turbinj}

Turbulence in the gas is continuously driven. The model used for turbulent driving 
is a generalization of the Ornstein-Uhlenbeck process developed and used by several 
authors \citep{eswaranExaminationForcingDirect1988,
schmidtNumericalDissipationBottleneck2006,
schmidtNumericalSimulationsCompressively2009,
federrathComparingStatisticsInterstellar2010}. It has been described in length 
in these articles, but we reproduce the description here for the sake of 
self-sufficiency and to introduce useful notation.

The force is computed in Fourier space and then applied to the gas. The evolution 
of the Fourier modes $\bm{\hat{f}}$ of the force is obtained via the resolution 
of the following differential equation:
\begin{equation}
\label{eq:ed_fourier}
    \mathrm{d}\bm{\hat{f}}(\bm{k}, t) = - \bm{\hat{f}}(\bm{k}, t)\dfrac{\mathrm{d}t}{T_\mathrm{driv}} 
    + F_0(\bm{k})\bm{P_\chi}\left(\bm{k}\right) \mathrm{d}\bm{W}_t.
\end{equation}
In this equation, $\mathrm{d}t$ is the timestep for integration and $T_\mathrm{driv}$ is 
the autocorrelation timescale. In our simulations, 
$T_\mathrm{driv}\approx L_\mathrm{box}/ (2 \sigma)$\footnote{The value of $T_\mathrm{driv}$ was set using the relation \eqref{eq:Mach_frms}. }. $\mathrm{d}\bm{W_t}$ is a small vector 
randomly chosen following the Wiener process, as described in 
\cite{schmidtNumericalSimulationsCompressively2009}. The power spectrum $F_0$ of 
the turbulent driving is:
\begin{equation}
    \label{eq:F0}
    F_0(\bm{k}) = 
    \begin{cases} 
    1 - \left(\dfrac{\bm{k}}{2\pi} - 2\right)^2\text{ if } 
    1 < \dfrac{\vert k \vert}{2\pi} < 3 \\
    0 \text{ if not.}
    \end{cases}
\end{equation}
The projection operator $\bm{P_\chi}$ is a weighted sum of the components of 
the Helmholtz decomposition of compressive versus solenoidal modes:
\begin{equation}
 \label{eq:projection}
    \bm{P_\chi}(\bm{k}) =  (1 - \chi) \bm{P}^{\perp}(\bm{k}) + 
    \chi \bm{P}^{\parallel}(\bm{k}) \;,
\end{equation}
with $\bm{P}^{\perp}$ and $\bm{P}^{\parallel}$ the projection operators 
respectively perpendicular and parallel to $\bm{k}$ 
\citep{federrathComparingStatisticsInterstellar2010}, and $\chi$ the compressive 
driving fraction. We tested three different cases: $\chi = 1$ (purely 
compressive), $\chi = 0$ (purely solenoidal), and $\chi = 0.5$ (natural mix 
between compressive and solenoidal modes). This compressive driving fraction 
applies only to the driving and is different from the compressive ratio measured 
in the velocity field. 
The forcing field $\bm{f}(\bm{x}, t)$ is then computed from the Fourier 
transform:
\begin{equation}
\label{eq:injection}
\bm{f}(\bm{x}, t) = g(\chi) f_{\mathrm{rms}}  \int\bm{\hat{f}}(\bm{k}, t) 
e^{i\bm{k}\cdot x} d^3\bm{k}\;.
\end{equation}
The parameter $f_{\mathrm{rms}}$ is directly linked to the power injected by 
the turbulent force into the simulation. The $g(\chi)$ factor is an empirical 
correction so that the resulting time-averaged root-mean-square of the power of 
the Fourier modes is equal to $f_{\mathrm{rms}}$, independently of the compressive 
fraction $\chi$.
We explore values of $f_{\mathrm{rms}}$ between 50 and 50000. 

\subsection{Self-gravity and star formation}
\label{subsec:gravity}

For each of our turbulent box simulation, gravity is added after two crossing times of the turbulence to see the outcome in terms of SFR.
The initial density and velocity field for these simulations are extracted 
from their non-gravitational counterparts at $2 T_\mathrm{driv}$, with 
$T_\mathrm{driv}$ the autocorrelation time of the turbulent driving introduced 
in Section \ref{subsec:turbinj}. By doing so, we ensure that the turbulence 
is fully developed once gravity is turned on.

The gravitational potential is computed via a multi-grid Poisson solver.
Star formation is tracked via sink particles \citep[e.g.][]{Krumholz04,
bleulerMoreRealisticSink2014}. Sinks are created when the gas density is 
above a threshold $\rho_{\mathrm{sink}}~=~10^3~\mu~m_p~ \mathrm{cm}^{-3}$ and subsequently accrete the gas over this threshold within there accretion radius of 4 cells (about 8 pc for the L1000\_N512 simulations).

\section{Measuring the density distribution: simulations without gravity}
\label{sec:nograv}

Here we present the results of the first step of the simulations, without self-gravity. We use them to study the statistical 
properties of the flow through the density PDF as well as the density power spectrum 
with particular attention to their dependence on the driving parameters. This 
is crucial to predict the SFR.

\subsection{Column density maps}
\label{subsec:coldens_nograv}

To get a first qualitative view of how turbulence shapes the density distribution, 
Fig.~\ref{fig:coldens_nograv} features column density maps integrated along the 
$z$-axis. The first striking observation is that stronger turbulence (increasing Mach number from the top to the bottom panels) generates 
more dense gas and sharper density fluctuations. The compressive fraction of the 
turbulent driving also significantly changes the outcome. When the driving is 
more compressive, denser regions are more extended. The contrast is sharper, and 
both high-density and low-density regions are larger and more contiguous. 
Qualitatively, the spatial structure of the density field in the compressible 
case should favour higher SFR compared to the solenoidal one, where the flow is 
more organised in smaller density structures. In what follows, we quantify 
these features as they play a role regarding the SFR.

\subsection{Mach number}
\begin{figure}
    \centering
    \includegraphics[width=\linewidth]{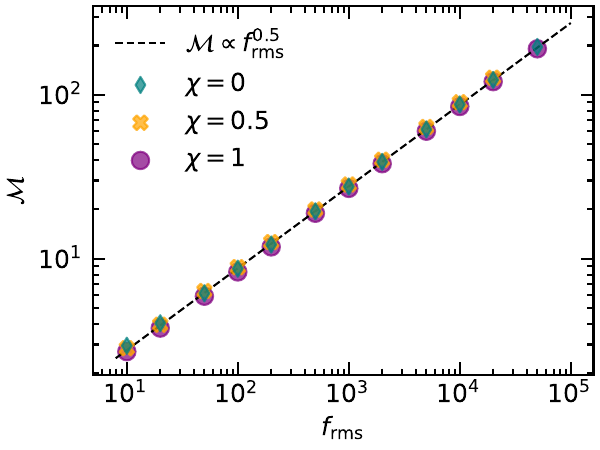}
    \caption{Mach number as a function of the RMS acceleration $f_\mathrm{rms}$ of the driving for the group L1000\_N512.}
    \label{fig:frmsmach}
\end{figure}

The Mach number ${\cal M}$ is one of the fundamental numbers used to characterise astrophysical flows. 
It is important to precisely quantify it for the various simulations.
Particularly, the forcing parameter $f_\mathrm{rms}$, which controls the amplitude of 
the forcing, has a direct influence on it.

The Mach number is given by
\begin{equation}
    \mathcal{M} = \dfrac{\sigma}{c_\mathrm{s}} \;,
\end{equation}
where $\sigma$ is the 3D mass-averaged velocity dispersion computed in the 
whole simulation domain and $c_\mathrm{s}$ is the speed of sound. At $10$~K, 
$c_\mathrm{s} = 0.2883 \kms$. 
The Mach number is a direct function of the strength 
of the turbulent driving and the size of the box, as shown by 
Fig.~\ref{fig:frmsmach}, with the relation:
\begin{equation}
    \label{eq:Mach_frms}
    \mathcal{M} = 0.90 f_\mathrm{rms}^{0.50}.
\end{equation}
We see in particular that the compressibility parameter\footnote{as expected from the normalization factor $g(\chi)$ of equation \eqref{eq:injection}}, $\chi$, does not have a 
significant influence on ${\cal M}$. Our simulations explore a large range of 
Mach numbers, from a few up to values larger than 100.

\subsection{Density PDF}

\begin{figure*}
    \centering
    \includegraphics[width=\textwidth]{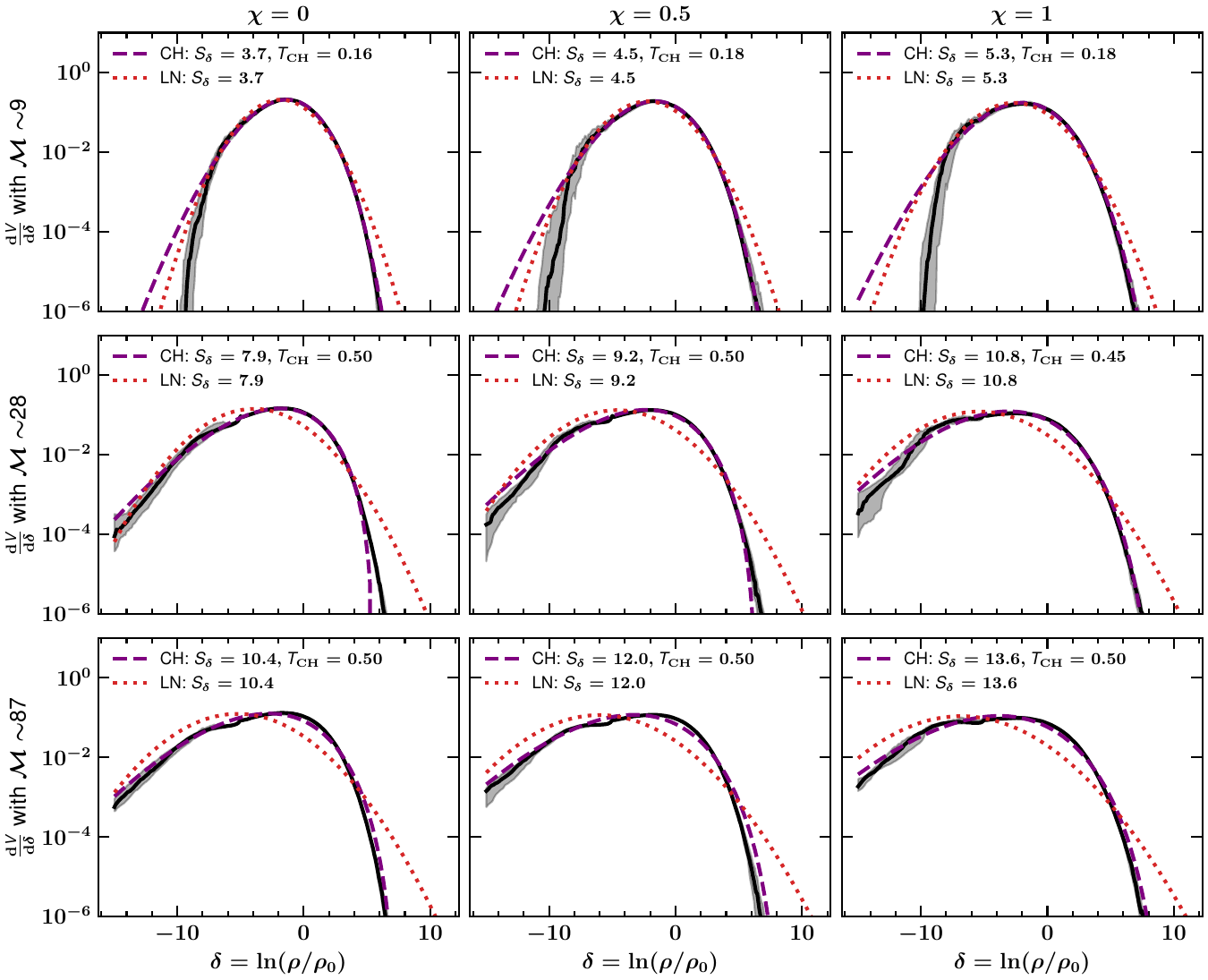}
    \caption{Time-averaged PDF of the natural logarithm of the density for the simulations 
    of the group L1000 without gravity. The black solid line is the time-averaged PDF 
    between 2 and 8 crossing times of the turbulence while the grey area depicts 
    its variation, containing 68\% of the data points. The purple dash line 
    is a fit (on the $T_CH$ parameter) of the Castaing-Hopkins PDF \citep{hopkinsModelNonlognormalDensity2013} 
    and the red dotted line is the log-normal PDF.}
    \label{fig:pdf_nograv}
\end{figure*}

\begin{figure}
\centering
    \includegraphics[width=0.4\textwidth]{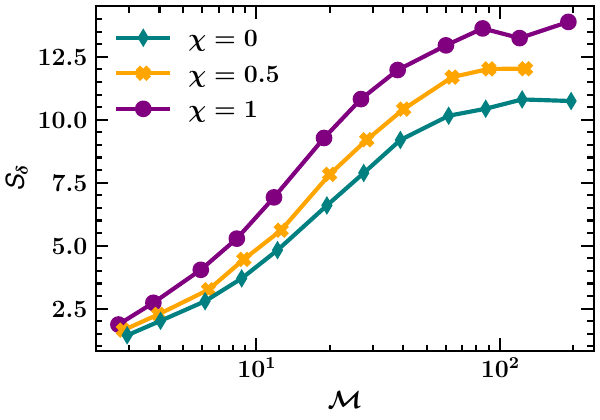}
    \caption{Volume-weighted variance of the natural logarithm of density as a function 
    of the Mach number and compressibility of the driving for the group L1000\_512}
    \label{fig:turbox_Slogrho}
\end{figure}

\begin{figure}
\centering
    \includegraphics[width=0.4\textwidth]{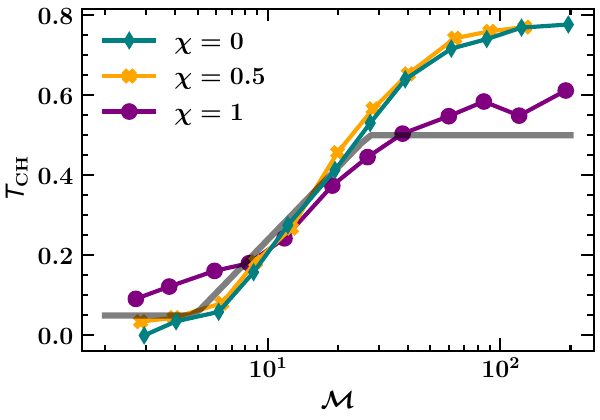}
    \caption{Fitted $T_\mathrm{CH}$ parameter of the Castaing-Hopkins PDF \citep{hopkinsModelNonlognormalDensity2013} as a function
    of the Mach number and compressibility of the driving for the group L1000\_512.
    Here we allowed the fit to explore the full [0-1] range of possible values, while in Fig. \ref{fig:pdf_nograv} we imposed $T_\mathrm{CH} < 0.5$ to avoid nonphysical drop of the PDF at high density. 
    Grey solid line is a rough approximation used in \citetalias{hennebelleInefficientStarFormation2024} and in Fig. \ref{fig:comparison_with_models}. }
    \label{fig:turbox_T_CH}
\end{figure}

\label{subsec:pdf}

The PDF of the gas density (or natural logarithm of density) in supersonic isothermal flows has been studied 
extensively over the years, mainly through numerical simulations. 
This one-point statistics measure gives important information about the amount of dense and less dense gas.
In Fig.~\ref{fig:pdf_nograv}, we show the time-averaged and volume-weighted PDF of the natural logarithm of density $\delta$ for a selection 
of simulations from the group L1000\_N512. Each PDF was computed by averaging the 
value of the distribution function in each ln-density bin from two to four 
autocorrelation times\footnote{which, we recall, is equal to one crossing time of the turbulence at the scale of the simulation box.} $T_\mathrm{driv}$. This results in 20 snapshots being stacked together.
The width of the PDF increases with the Mach number and the compressibilty of the driving.
This is quantified in Fig.~\ref{fig:turbox_Slogrho}, which portrays the variance of $\ln \rho$, 
$S_{\delta}$, as a function of Mach number for three values of the compressibility $\chi$. 
As expected, $S_{\delta}$ increases with ${\cal M}$. For ${\cal M}$ below 10, 
it is lower than 4 and it seemingly reaches a maximum around ${\cal M} \simeq 100$. 
The value of this maximum depends on the compressibility parameter, $\chi$.
It is about 10 for $\chi=0$ and 13 for $\chi=1$. 
The plateau at high Mach number is probably a resolution effect, as can be seen is Fig. \ref{fig:conv}. At a lower resolution of $256^3$, the plateau appear sooner at $\mathcal{M} \sim 30$, while at a higher resolution of $1024^3$ no plateau is seen.

Several empirical and theoretical formula have been used for the volume-weighted density PDF. 
\citet{vazquez-semadeniHierarchicalStructureNearly1994} and 
\citet{nordlund1999} proposed the most widely used log-normal (LN) distribution:
\begin{equation}
\label{eq:pdf_formula_log-normal}
\dfrac{\mathrm{d}V}{\mathrm{d}\delta}_\mathrm{LN} = \dfrac{1}{\sqrt{2 \pi S_{\delta }}} \exp\left(\dfrac{-\left(\delta - \delta_0\right)^2}{2 S_{\delta }}\right),
\end{equation}
where $\delta = \ln\left(\rho / \rho_0 \right)$ with $\rho_0$ is the mean density, $\delta_0 = S_{\delta }/ 2$, and $S_{\delta }$ the volume-weighted variance of the natural logarithm of the normalized density 
$\delta= \ln \left(\rho/\rho_0 \right)$.

It has been pointed by several studies that the PDF is no longer log-normal at high Mach number \citep{federrathComparingStatisticsInterstellar2010,hopkinsModelNonlognormalDensity2013,squireDistributionDensitySupersonic2017,moczMarkovModelNonlognormal2019}. A useful distribution of the density PDF has been given by 
\citet{hopkinsModelNonlognormalDensity2013} following the expression obtained 
by \citet{castaing1996} to describe the velocity PDF of incompressible flows. 
The proposed functional form, hereby called Castaing-Hopkins (CH) PDF, is given by:
\begin{equation}
\label{eq:hopkins_pdf}
    \dfrac{\mathrm{d}V}{\mathrm{d}\delta}_\mathrm{CH}  = \begin{cases}
    \dfrac{1}{T_\mathrm{CH}}  \displaystyle\sum_{m=1}^{\infty}
    \dfrac{\lambda^m e^{-\lambda}}{m!} 
    \dfrac{u^{m-1} e^{-u}}{(m-1)!} & \text{ if }  u \geq 0,\\
    0 & \text{ if }  u < 0,
 \end{cases}
\end{equation}
where
\begin{align}
    \label{eq:u_and_lambda}
    u &= \dfrac{\lambda}{1 + T_\mathrm{CH}} - \dfrac{\ln \rho}{T_\mathrm{CH}}, \\
    \lambda &= \dfrac{S_\delta}{2 T_\mathrm{CH}^2},
\end{align}
and $T _{\mathrm{CH}}$ is a parameter that describes how much the function deviates from a log-normal.

\begin{figure*}[ht]
    \centering
    \includegraphics[width=\textwidth]{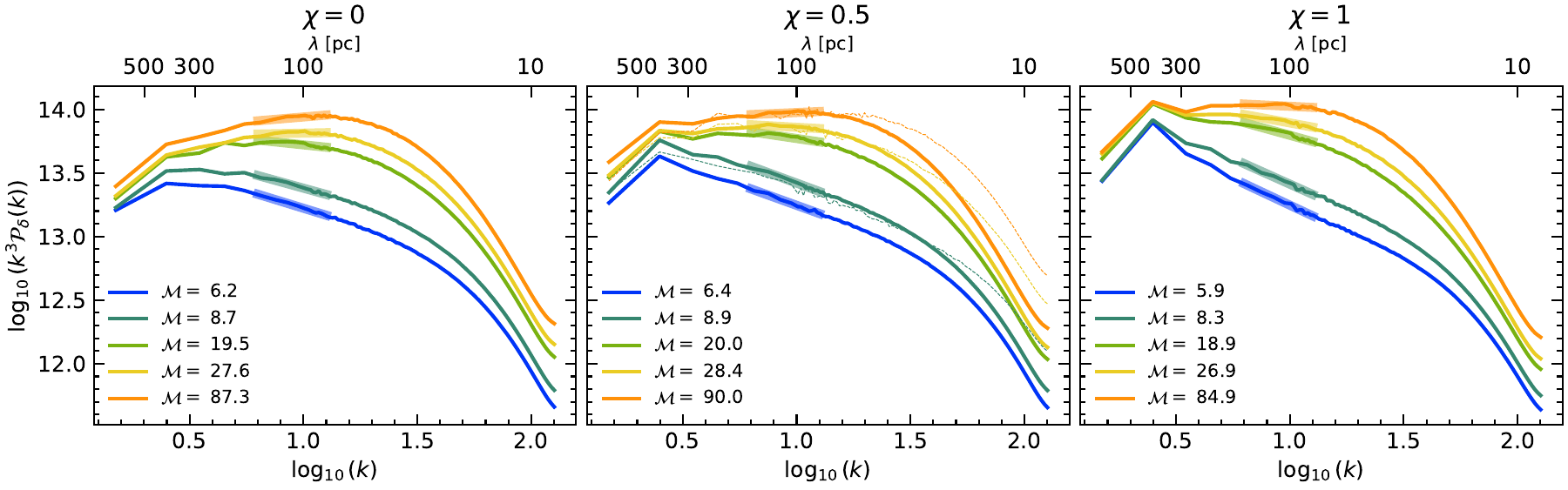}
    \caption{Power spectra of the natural logarithm of density for the simulation of the group L1000\_512 (solid) and L1000\_1024 (dotted lines in middle panel). The thicker range show the captured inertial range used to fit the slope and compute $\eta_d$ in Fig. \ref{fig:turbox_eta}. It is defined as the portion of the curve that is left unchanged when resolution is increased. 
    The displayed power spectra are compensated (multplied by $k^3$) such that a power-law slope of $-3$ is horizontal on the plot.
    }
    \label{fig:turbox_pspec}
\end{figure*} 

\begin{figure}
    \centering
    \includegraphics[width=0.4\textwidth]{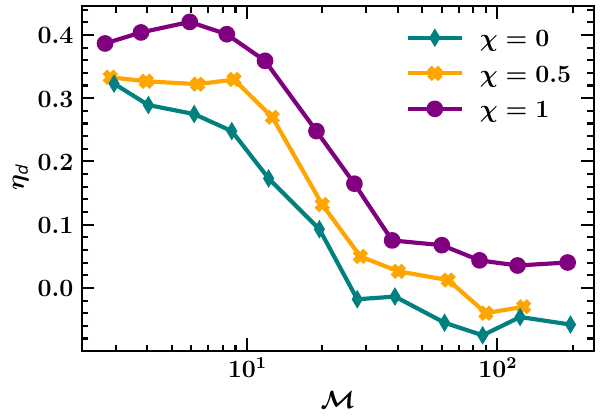}
    \caption{The $\eta_d$ parameter computed from the slope of the power spectrum of the natural logarithm of density (Fig. \ref{fig:turbox_pspec}, Eq. \eqref{eq:eta_d_s}) as a function of the Mach number and compressibility of the driving.}
    \label{fig:turbox_eta}
\end{figure}

In Fig.~\ref{fig:pdf_nograv}, the comparisons of our simulated data with a log-normal and a Castaing-Hopkins PDF are also given. 
The variance  $S_{\delta}$ of $\delta$  (see Fig.~\ref{fig:turbox_Slogrho}) is computed directly from the time-averaged density distribution measured in the simulations.
The parameter of the Castaing-Hopkins PDF $T_{\mathrm{CH}}$ was fitted on 
the data (see Fig.~\ref{fig:turbox_T_CH}). The value of $T_{\mathrm{CH}}$ 
is capped to 0.5 to avoid a sharp drop-off of the PDF at high density.  
We also fitted the parameters of the log-normal as a test, and that did not 
give qualitatively better results.

Whereas the agreement with log-normal PDF is not catastrophic for  ${\cal M} \approx 9$, the 
density PDF for higher Mach values is clearly non-log-normal as written above and anticipated 
by \cite{federrathComparingStatisticsInterstellar2010}. 
On the other hand, the Castaing-Hopkins density 
PDF provides a much better fit, as already found by 
\citet{hopkinsModelNonlognormalDensity2013}, who has confronted Eq.~(\ref{eq:hopkins_pdf}) 
to a large suite of numerical simulations. 
In particular, it is seen that at high Mach number, the shape of the PDF is 
less and less symmetrical with respect to the mean density, with the low 
density distribution presenting a more prominent shape. 
Interestingly, Eq.~(\ref{eq:hopkins_pdf}) is believed to describe a log-Poisson 
turbulent cascade. In the context of incompressible flows, this intermittent 
model leads to structure function coefficients that are in excellent 
agreement with laboratory experiments \citep{castaing1996}. 
Noticeably, it is found that the log-normal PDF predicts significantly more 
dense gas than what is obtained from numerical simulations, implying that at 
high Mach numbers, models using the log-normal PDF may lead to over-estimated SFR.

Figure ~\ref{fig:turbox_T_CH} shows the result of the fit of the parameter $T_\mathrm{CH}$, when the full range [0-1] range of possible values is explored. In practice, value of  $T_\mathrm{CH} > 0.5$ unrealistically cut off the higher densities while still very well fitting the rest of the PDF.
The parameter $T_\mathrm{CH}$ increases with Mach, and the slope is steeper when the compressibility of the driving is lower. 
The grey line depicts an approximation of the value of $T_\mathrm{CH}$ as a function of the Mach that can be used in analytical models, such as the one we present in \citetalias{hennebelleInefficientStarFormation2024}. 
\begin{equation}
\label{eq:T_fit}
    T_\mathrm{CH} = \min(\max( 0.6 \log_{10}(\mathcal{M} / 4, 0.1), 0.5).
\end{equation}
For simplicity, we use a compressibility-independent approximation.

\subsection{Power spectrum of the density}
\label{subsec:ps}

In Paper I, we emphasize the importance of correctly considering the spatial 
correlation of the density distribution. The density PDF lacks information on how 
the dense gas is spatially distributed. However, the SFR should depend on it to 
some extent. From Fig.~\ref{fig:coldens_nograv}, it is evident that different 
drivings create structures with a different size distribution.
We can estimate whether the dense gas is more preferentially found in small or 
large structures and, more generally, describe the spatial density distribution 
by looking at the power spectrum of the natural logarithm of density. We consider the 
natural logarithm of the density to be the relevant quantity because the density PDF, a 
fundamental quantity regarding the SFR, is accurately characterised using 
$\ln \rho$. In particular, the scale dependence of the natural logarithm of density variance plays a 
role in the analytical model presented in Paper~I and directly relies on the 
power spectrum of $\ln \rho$.

Figure~\ref{fig:turbox_pspec} shows spectra of $\ln \rho$ 
for a selection of simulations. Each spectrum reveals three distinct regions: 
the injection scales with $1 \leq k \leq 3$ (that is $0 \leq \log_{10} k \leq 0.48$), 
the inertial range depicted with the thick bar (for our resolution, $0.6 \leq 
\log_{10} k \leq 1.1$), and the dissipation range ($\log_{10} k \geq 1.5$). The inertial 
range was determined through a resolution study (see 
Section~\ref{subsec:resolution_pspec}).
The power-law exponent $n_d$ of the power spectrum is of high interest 
because it provides insight into the variation of the natural logarithm of the density PDF over 
scale. It is linked to the $\eta_d$ parameter introduced in \citetalias{hennebelleInefficientStarFormation2024} via the 
relation:
\begin{equation}
\label{eq:eta_d_s}
    \eta_d = - \dfrac{3 + n_d}{2}\;.
\end{equation}
In our simulations, we find slopes $n_d$ in the inertial range between $-4$ ($\eta_d = 0.5)$, which 
we will call steep, and $-3$, which we will call flattened. When 
the power spectrum is steep, most of the power is in the large scale, meaning 
that the dense gas is contained within a few big structures. On the other hand, 
a flattened power spectrum indicates more power in the small structures, 
meaning that the dense gas is fragmented into many small structures.

The dependence of the slope $n_d$ on the driving parameter is already quite 
clear from Fig.~\ref{fig:turbox_pspec}: the ln-density power-spectrum flattens 
with increasing Mach number and steepens with increasing compressibility. This is 
even clearer in Fig.~\ref{fig:turbox_eta}, where the value of the parameter 
$\eta_d$ computed from Eq.~\ref{eq:eta_d_s} is plotted as a function of the 
Mach number for different compressibility fractions.
For solenoidal and mixed drivings, the parameter $\eta_d$ goes from 
$\approx 0.3$ below $\mathcal{M} = 10$ and quickly decreases towards $\approx 
0$ for $\mathcal{M} = 30$, after which it saturates. The evolution is similar 
for compressive driving, but with higher values of $\eta_d$, between $0.4$ and 
$0.1$.

Physically, as the Mach number increases, the flow organises into denser but 
smaller scale structures (the power spectrum becomes flatter). An important effect of the compressibility parameter of the driving, 
$\chi$, is not only to modify the density PDF, making it broader for 
$\chi=1$ than $\chi=0$, but $\chi$ also slightly influences the typical 
size of dense structures. In particular, more compressible flows remain 
spatially more coherent than solenoidal ones (and thus the power spectrum for compressible flow is steeper).

\subsection{Replenishment time}
\label{subsec:repl_time}

\begin{figure}
    \centering
    \includegraphics[width=\linewidth]{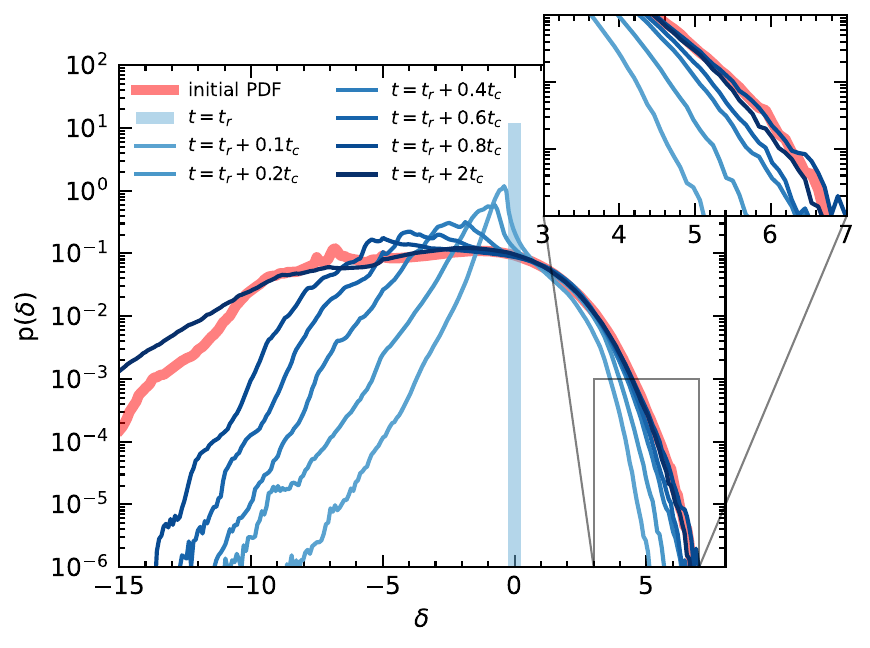}
    \caption{Evolution of the density PDF ($\delta = ln(\rho/\rho_0)$) after the density was reset at $\rho_0$ from a snapshot of the simulation L1000\_N512\_comp0.5\_frms1e4. The red line is the PDF just before the reset, and the blue lines corresponds to different times given in units of the crossing time of the turbulence $t_c$. We call $t_r := 2 t_c$ the time at which the density has been reset.}
    \label{fig:turbox_repl_time}
\end{figure}

\begin{figure*}
    \centering
    \includegraphics[width=0.83\textwidth]{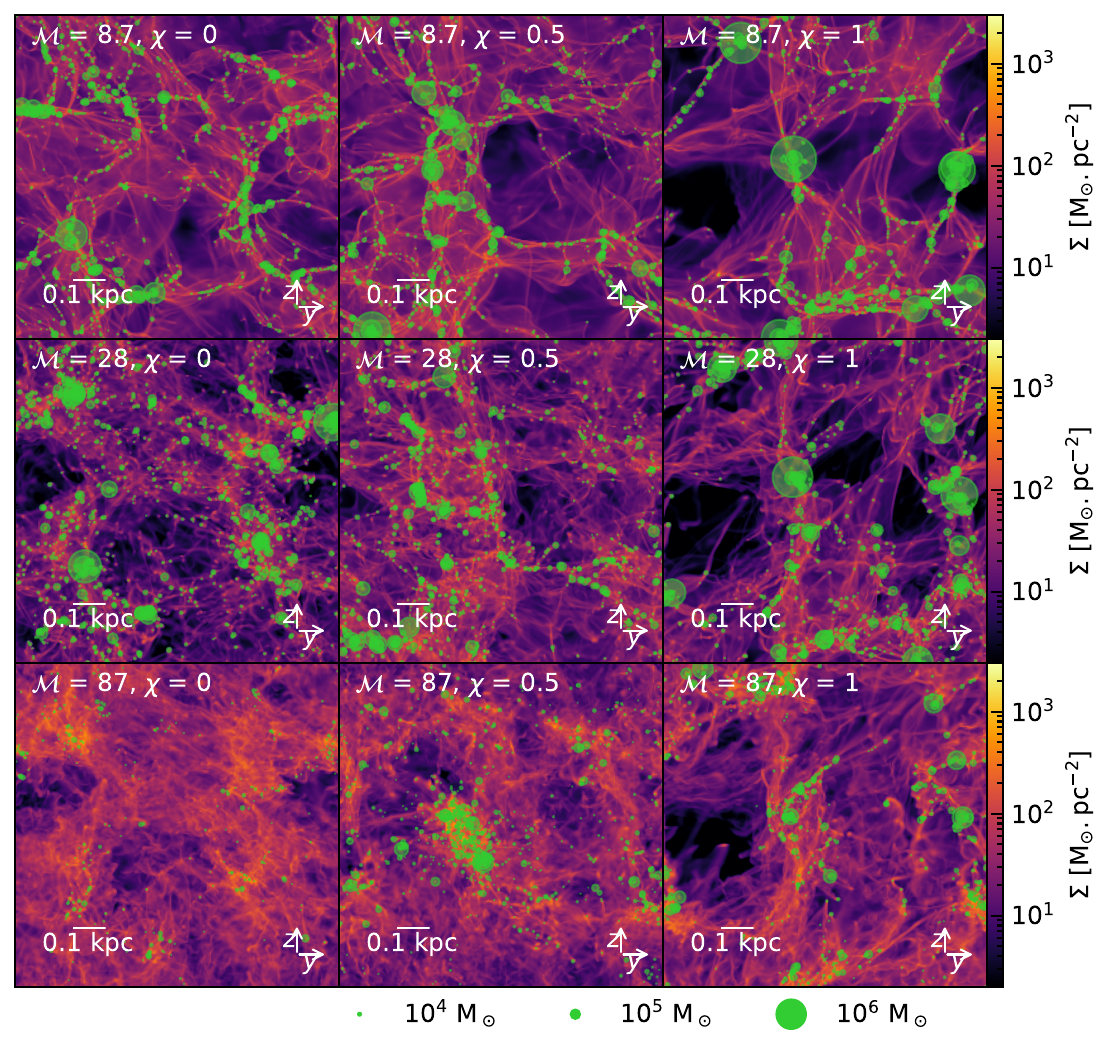}
    \caption{Column density for the simulations with gravity at $t= 0.2 t_\mathrm{ff}(\rho_0)$ after the activation of gravity. The green circles represent the sink particles with their radius proportional to the decimal logarithm of their mass.}
    \label{fig:coldens_grav}
\end{figure*}

The replenishment time, the time needed for turbulent flow to rebuild the 
distribution of gravitationally unstable structures once they have collapsed, 
plays a crucial role in determining the SFR. In \citetalias{hennebelleInefficientStarFormation2024}, we introduce a new 
analytical method to estimate it and demonstrate that it is the primary control 
time of star formation in the high-Mach regime. 

Measuring the replenishment time in simulations is not straightforward. In 
simulations with gravity, where dense structures can collapse, the effects of 
turbulence and gravity are intermingled. 
We illustrate the time needed by turbulence to replenish dense structures with a simple numerical experiment. After two crossing time at the scale of the box $t_c := \sigma / L_\mathrm{box}$, that is when the turbulence is fully developed, we 
reset the simulation to the initial density, while retaining the developed 
velocity field. This process does not change the mass in the simulation, but just redistribute it back to its initial state.
We call the reset time $t_r := 2 t_c$.
We then measure the time required to rebuild different dense 
structures at various scales, essentially assessing how long it takes to 
rebuild the density PDF.
Figure \ref{fig:turbox_repl_time} shows the density PDF at several times measured in unit of crossing time $t_c$ after the density has been reset to its initial value for the simulation L1000\_N512\_comp0.5\_frms1e4.
It takes about two crossing times of the turbulence to fully rebuild the PDF. 
Crucially, how long it takes to rebuild structures highly depend on their density. 

Focusing on the over-dense part of the plot ($\delta > 0$), we remark that the stronger the over-densities is, the longer it takes to rebuild it.
Small over-densities ($0 < \delta < 3$) are rebuilt rather quickly, in less than $0.2 t_c$. 
On the contrary, stronger over-densities ($\delta > 5$) which also correspond to smaller structures requires almost a crossing time ($\sim 0.8 t_c$) to be rebuilt.
We remind that here $t_c$ is the crossing time at the scale of the box $L_\mathrm{box}$, which is much longer than the crossing time at the scale of such dense structures, used to estimate the replenishment time in for previous studies \citep[e.g][]{hennebelle2013}. 
For completeness, we note that the under-dense part ($\delta < 0$) requires even longer time to be fully rebuilt, with the very low density zone ($\delta < -12$) appearing between one and two crossing times. This part of the PDF is less crucial for the star formation and undergo more fluctuations, as can be seen in Fig.~\ref{fig:pdf_nograv}.

\section{Measuring the SFR: simulations with gravity}
\label{sec:grav}

After running the simulations without gravity for two crossing times, self-gravity is 
activated, initiating gravitational collapse. As previously explained, sink 
particles are introduced when the local Jeans length is not resolved by at 
least ten points.

Figure~\ref{fig:coldens_grav} displays the column density of the simulations 
one fifth of a free-fall time after gravity has been switched on.
In simulations with weak turbulent 
driving (top row), prominent self-gravitating filamentary structures develop 
almost immediately. Further collapse occurs along the filaments, resulting in 
aligned series of sink particles. 
These simulations exhibit a distinct appearance compared to simulations without gravity, as presented in Fig.~\ref{fig:coldens_nograv}. 
On the other hand, the filamentary network is less distinct when the driving is 
stronger, and fewer sink particles are formed. These simulations visually 
resemble their non-self-gravitating counterparts.

\begin{figure*}[ht]
    \centering
    \includegraphics[width=0.85 \textwidth]{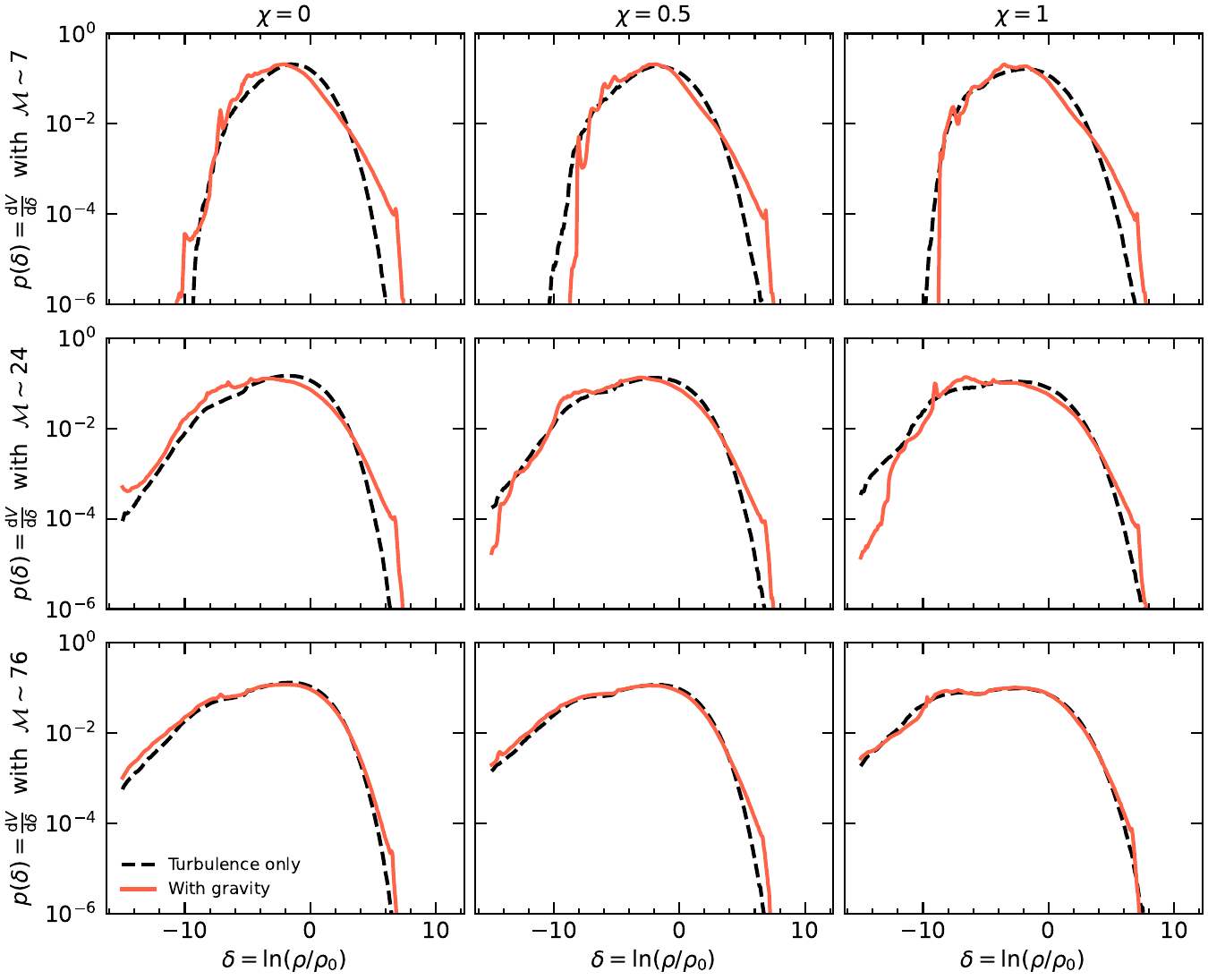}
    \caption{Density PDF for the simulations with gravity at $t= 0.2 t_\mathrm{ff}(\rho_0)$ after the activation of gravity. }
    \label{fig:pdf_grav}
\end{figure*}

\subsection{Evolution of the PDF}
\label{subsec:pdf_grav}

Figure~\ref{fig:pdf_grav} portrays the density PDF, one fifth of a free-fall time after gravity was switched on.
Compared to the density PDFs obtained in the absence of self-gravity, the PDF are broader and there 
is more dense gas especially for low Mach numbers.
We note in particular the apparition of a power law tail, compatible with $dN/d \delta \propto \rho^{-1.5}$, except for the highest Mach numbers
as reported in earlier works \citep[e.g.][]{kritsukDensityDistributionStarforming2011,HF12}.
Let us remind that a power law $dN/d \ln \rho \propto \rho^{-1.5}$ can be interpreted as a simple consequence of freefall. 
Indeed, in the region of a spherical infalling envelope which is dominated by the gravity of the gas, a dependence $\rho \propto r^{-2}$ is expected, 
where $r$ is the distance to the collapse center. Since the number of fluid particles, $d N$, located between $r$ and $r+dr$ is such that 
$d N \propto r^2 dr$ and thus $d N \propto \rho^{-1 -3/2} d \rho$, which leads to $d N / d\delta  \propto \rho ^{-3/2}$.

We stress that beyond the power law  index  of the PDF, which develops in the presence of self-gravity, a global understanding of the PDF
generated by gravo-turbulence is still lacking. In particular, what fraction of the mass would be contained in the power law part and 
at which density the transition between the turbulent and the self-gravitating PDF does occur remains to be understood.

\subsection{Star formation rate}

\begin{figure}
    \centering
    \includegraphics[width = \linewidth]{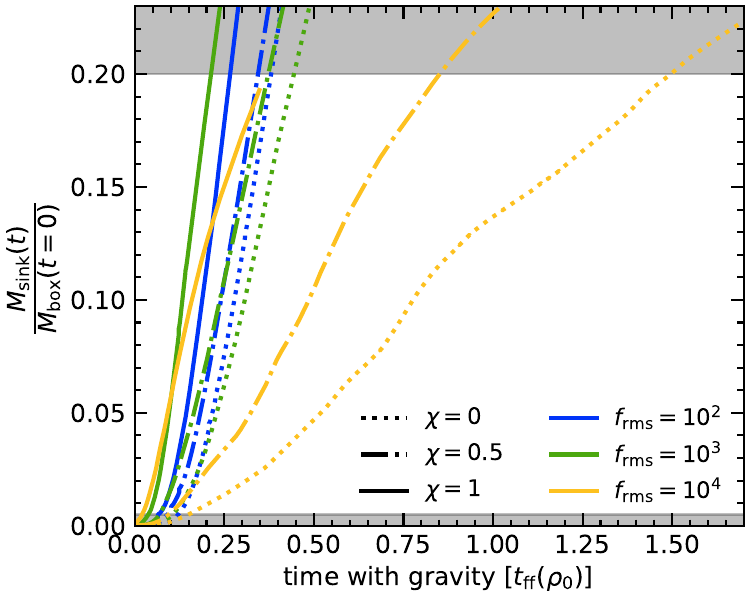}
    \caption{Fraction of the total initial mass which has been accreted in the sinks as a function of time since the activation of the gravity, in units of the freefall time at the density $\rho_0$. The SFR is computed between the two shaded areas.}
    \label{fig:mass_sink}
\end{figure}

\begin{figure}
    \centering
    \includegraphics[width=\linewidth]{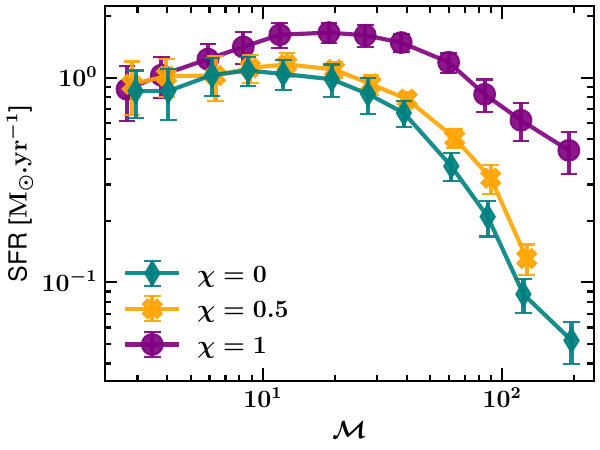}
    \caption{SFR as a function of the turbulent Mach number. The Mach number is computed from the corresponding simulations without gravity.}
    \label{fig:sfrmach}
\end{figure}

To estimate the star formation rate (SFR), we measured the time it takes for 20\% 
of the gas mass to be accreted into sinks. Specifically, we define \(t_{0.5\%}\) to be the 
time needed to convert 0.5\% of the gas into sinks, and 
\(t_{20\%}\) he time needed to convert   20\%.
The SFR is then computed the following way:
\begin{equation}
    \mathrm{SFR} = \dfrac{M_{\mathrm{sink}}(t_{20\%}) - M_{\mathrm{sink}}(t_{0.5\%})}
    {t_{20\%} - t_{0.5\%}} = \dfrac{0.195 M_{\mathrm{box}}(t=0)}{t_{20\%} - 
    t_{0.5\%}},
\end{equation}
where \(M_{\mathrm{sink}}(t)\) and \(M_{\mathrm{box}}(t)\) are respectively the mass in 
the sinks and in the box at time \(t\) 
(\(M_{\mathrm{box}}(t) = M_{\mathrm{box}}(t=0) - M_{\mathrm{sink}}(t)\)). Note that we 
do not estimate the SFR starting from \(t=0\) since, as explained above, we start 
from a simulation that does not include self-gravity, and therefore the earliest 
times are not representative. On the other hand, the choice of \(t_{20\%}\) is not 
critical. This can be easily seen in Fig.~\ref{fig:mass_sink}, which shows the 
mass accreted in sinks as a function of time, expressed as a fraction of the 
total mass of the box. The SFR is the slope of the featured curves. It appears 
that the SFR is very steady, and the precise value of when it is measured does 
not matter much for the end result.

Figure~\ref{fig:sfrmach} portrays the SFR as a function of Mach number. 
for three values of the driving compressive fraction \(\chi\), namely 0, 0.5, and 1. 
For Mach numbers below \(\simeq 20\), the SFR is relatively constant with Mach, being 
slightly larger (by about 50\% at \(\mathcal{M}=20\))
for \(\chi=1\) than for \(\chi=0\) and 0.5. For larger Mach numbers, the SFR decreases steeply with \(\mathcal{M}\). 
For \(\chi=0\) and 0.5, the SFR at \(\mathcal{M} \simeq 150\) has decreased by more than a factor of 10 compared to its value
at \(\mathcal{M} \simeq 20\). For \(\chi=1\), the SFR decrease remains more limited, and globally the SFR is significantly 
larger than for \(\chi=0\) and 0.5, particularly for \(\mathcal{M} > 20\). 

We thus identify two regimes: at low Mach numbers the SFR is fairly constant or slightly increasing with $\mathcal{M}$, while at high $\mathcal{M}$ there is a decrease. For our injection scales, the transition between the two regimes  occurs
around $\mathcal{M}$ $\simeq$ 20.
This is in agreement with the theoretical prediction of \citetalias{hennebelleInefficientStarFormation2024}.

\subsection{Comparisons between models and simulations}
\label{subsec:comparison_models}

\begin{figure}
    \centering
    \includegraphics[width=0.98 \linewidth]{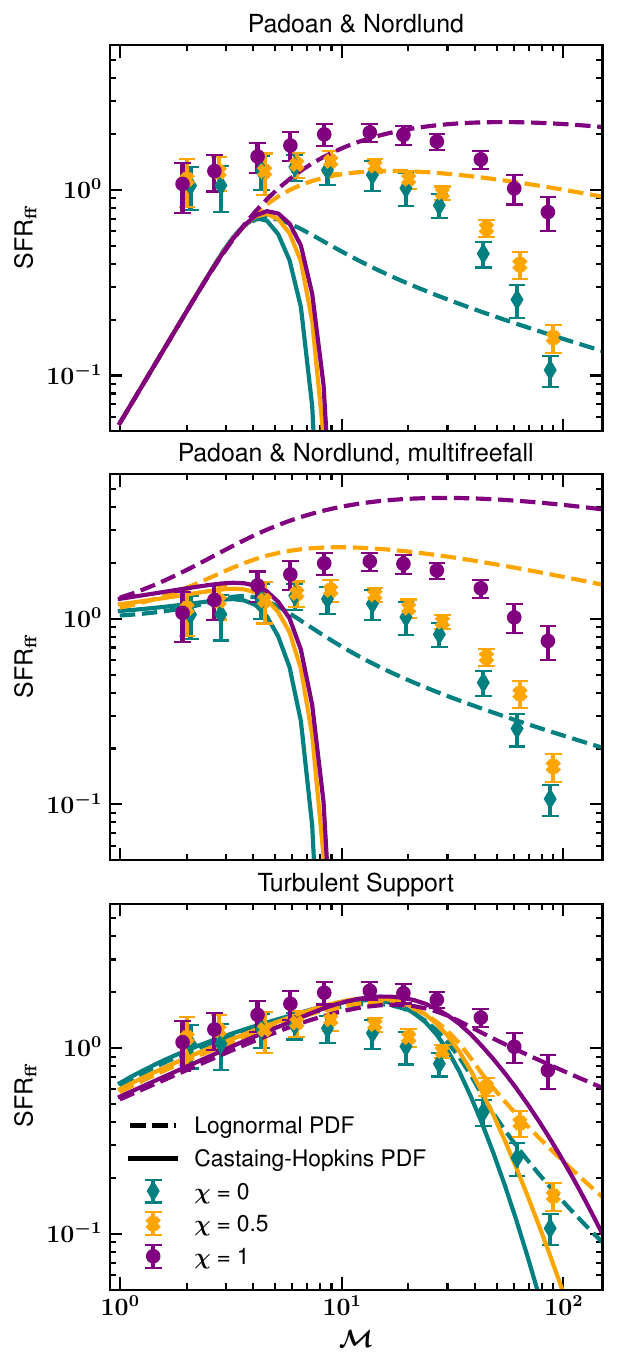}
    \caption{Comparison of the normalized SFR from the model of \cite{padoanStarFormationRate2011} (top), the multifreefall version of it (middle), the TS model of \citetalias{hennebelleInefficientStarFormation2024} (bottom) and the results from the simulations of this work (markers) for compressive ($\chi = 1$), mixed ($\chi = 0.5$), and solenoidal ($\chi = 0$) driving. Each model is used with a log-normal density PDF (dashed lines) and a Castaing-Hopkings PDF (solid lines).}
    \label{fig:comparison_with_models}
\end{figure}

Several gravo-turbulent models for the SFR has been developed over time \citep{padoanStarFormationRate2011,padoanSimpleLawStar2012,federrathStarFormationRate2012,hennebelleAnalyticalStarFormation2011,burkhartStarFormationRate2018}.
A review of the different models is available in \citetalias{hennebelleInefficientStarFormation2024}.
We built our simulations to be as close as possible to the idealized turbulent ISM of the gravo-turbulent model, assuming that the turbulence shapes gravitationally unstable clouds that further collapse. 
This way, we are able to perform a direct comparison with the theoretical models, with the main limitation that our resolution is limited (see Section~\ref{sec:caveats}).

In Fig. \ref{fig:comparison_with_models}, we compare the SFR measured in our simulations with the predictions of \cite{padoanStarFormationRate2011} (top panel) the adapted multifreefall version of it  presented in \cite{hennebelleAnalyticalStarFormation2011} (middle panel) and the new TS model presented in \citetalias{hennebelleInefficientStarFormation2024}, using the Castaing-Hopkins formulation of the PDF. 
For this comparison, we used models with $n_c = 1.5$ and an injection scale of 333 pc, with a cuting parameter $y_\mathrm{cut} = 0.5$.
For the TS model, we used $\eta_d = 0.4$ for $\chi =1$ and $\eta_d = 0.2$ for $\chi = 0$ or $0.5$. 
These values are in the range of values found in Fig.~\ref{fig:turbox_eta}. 
A more precise parametrisation of $\eta_d$ including the Mach dependence would be more satisfactory. 
Unfortunately, as we discuss in Section \ref{sec:caveats} and in the appendices, the fact that we measure $\eta_d$ in a small range of inertia makes it difficult to reliably infer the distribution of the gas at all scales.
The value of $T_\mathrm{CH}$ is the parametrisation of Eq.~\eqref{eq:T_fit}.
The plotted value of the SFR is normalized by the mass in the box divided by the free-fall time
\begin{equation}
    \mathrm{SFR}_\mathrm{ff} = \dfrac{t_\mathrm{ff}}{\rho_0 L_\mathrm{box}^3} \mathrm{SFR},
\end{equation}
where $t_\mathrm{ff}$ is the free-fall time at the mean density $\rho_0$:
\begin{equation}
    t_\mathrm{ff} = \sqrt{\dfrac{3 \pi}{32 G \rho_0}}.
\end{equation}

The model of \cite{padoanStarFormationRate2011}, used with the traditional log-normal PDF, fails in many points. 
First, it yields unrealistically low value of the SFR for Mach number $\mathcal{M} < 10$, with a very steep slope in that regime. 
The simulations show, on the contrary, rather high values of the SFR, and a shallow almost flat slope.  
In the high-Mach regime, it overestimates the effect of the compressibility, with a purely solenoidal driving yielding SFR one order of magnitude lower than the mixed driving, while the two show a similar behaviour in the simulations. 
More importantly, it fails to reproduce the huge drop in SFR for Mach numbers higher than $\mathcal{M} \sim 20$.
The multi-freefall model (middle panel) is in much better agreement with the simulations for  $\mathcal{M} < 10$ but fares no better in the high-Mach regimes.
Using a more realistic Castaing-Hopkins PDF only makes the prediction worse for both models, completely cutting off star formation at Mach numbers higher than 10.

The TS model is the only one to correctly reproduce the drop in the high-Mach regime while and also match the low Mach regime. 
The break of the slope falls down around $\mathcal{M} \sim 20$, as in the simulation. 
As explained in \citetalias{hennebelleInefficientStarFormation2024}, the position of the break is directly related to the scale of injection.
Here we used a scale of $333$ pc, which is the lowest scale over which the turbulent driving is acting.
However, the match is not perfect. 
First, the slope in the high-Mach regime is steeper in the TS model than in the simulation results when a Castaing-Hopkins PDF is used. This may be a consequence of the estimation of the width of the density PDF averaged at intermediate scales, a crucial ingredient of the TS model.
Indeed, because of numerical dissipation, the power spectra of Fig.~\ref{fig:turbox_pspec} cannot be characterized by a power-law over all scales.
This is further discussed in appendix \ref{sec:scale_evo}.
Surprisingly, the less realistic log-normal PDF predict a slope closer to the one measured in the simulations.

Second, the break is too abrupt, especially when the driving compressibility is low.
In the model, the break happens when the Mach number is such that all scales up to the injection scale are stabilized by turbulence. 
The injection scale is not as well-defined in the simulation, which may explain a smoother transition between the low-Mach and the high-Mach regimes.

\section{Caveats}
\label{sec:caveats}

In this study we try to test (and improve) the analytical model for SFR based on the gravo-turbulent theory.
It is first crucial to well understand the scope of this work: a full theory to predict the SFR would require to carefully model all the relevant processes involved at the various scales, including magnetic field and stellar feedback.
Instead, we focus on the effect of turbulence, defined as a structured way to drive random motion in the gas, and gravity. 
At the scales of hundred of parsecs, besides the important role of turbulence and gravity, the interstellar gas is also subject to the effect of the magnetic field and cosmic rays \citep{klessenPhysicalProcessesInterstellar2016, ibanez-mejiaGravityMagneticFields2022}.
At scales below, stellar feedback, radiation and thermodynamics can no longer be ignored \citep{gattoModellingSupernovadrivenISM2015,kimThreephaseInterstellarMedium2017,brucyLargescaleTurbulentDriving2020}. 
In analytical models, the effect of smaller scales is hidden behind a fudge factor that reduces the SFR. This is considered to be the efficiency for transforming small scale structure into actual stars. 
For our comparison purpose, a fudge factor is not needed as in the simulation we consider that all the gas over a given density is bound to form stars \citep{federrathStarFormationRate2012}.

Ideally, we would like to measure all the relevant quantities directly in the simulations and be able to use them as an input for the TS model of \citetalias{hennebelleInefficientStarFormation2024}.  
Globally, the quantities we rely on (variance of the ln-density, the SFR and the value of the $\eta_d$ parameter) are well converged at our fiducial resolution of $512^3$, as shown in appendix \ref{sec:conv}.
However, this is sometimes not enough. A good example would be the slope of the power spectrum of the ln-density, which plays an important role in the TS model.
Unfortunately, the resolution required to have an inertial range that cover a large enough range of scales is not reachable for such a parameter study. 
As a consequence, the parameter $\eta_d$ cannot be reliably used to probe the distribution of the gas at all scales, as explained in appendix \ref{sec:scale_evo}.

Another important caveat is the use of the Ornstein-Uhlenbeck process, which is an idealised way of driving the turbulence. 
Real turbulence driving source can produce velocity field with different properties. For instance,  \cite{gongImpactMagnetorotationalInstability2021} have shown in a different context that the magneto-rotational instability can produce velocity field more solenoidal than the most solenoidal Ornstein-Uhlenbeck driving ($\chi = 0$).

\section{Conclusions}
\label{sec:conclusion}

In this work we realize a series of (gravo-)turbulent isothermal simulations to probe how the interplay between turbulence and gravity shapes the SFR. We focus on environments with high velocity dispersion, notably found in $z \sim 1$ galaxies and the centre of local galaxies. 
Our conclusions are the following:
\begin{enumerate}
    \item For high-Mach number, the density PDF is no longer a log-normal. PDF models taking intermittency into account perfectly fits the PDF from simulations.
    \item Both the Mach number and the compressibility of the driving change how the dense gas is distributed. Weaker and more compressive driving will favor large-scale structures, while with a stronger and more solenoidal driving the dense gas is distributed in small-scale structures.
    \item Very dense structures requires more time to be created by the turbulence than less dense one. The time to build such dense structure is much higher that the crossing time at the scale of these structures.
    \item For isothermal gas, the SFR quickly drops when the Mach number is higher than a threshold that depends on the injection scale of the turbulence. With an injection scale between 300~pc and 1~kpc, this threshold is around $\mathcal{M} \sim 20$. This drop in SFR is not reproduced by classical, log-normal-based analytical models for the SFR in gravo-turbulent medium.
    \item Classical SFR analytical models fail to reproduce the inefficient high-Mach regime of star formation.
    \item While still not perfectly matching the exact shape of the SFR-$\mathcal{M}$ curve, the analytical model presented in \citetalias{hennebelleInefficientStarFormation2024} correctly reproduce the behaviour of the SFR in the high-Mach regime.
    This is because it more accurately models the density PDF, the density distribution of the gas and the time needed to replenish the unstable dense structures.
\end{enumerate}

\begin{acknowledgements}

NB thanks Gilles Chabrier, Marc-Antoine Misvilles-Deschênes, Jéremy Fensch, Cara Battersby, Guillaume Laibe, Pierre Dumond, and Elliot Lynch for interesting discussions about this work.
The authors acknowledge Interstellar Institute's program "II6" 
and the Paris-Saclay University's Institut Pascal for hosting discussions 
that nourished the development of the ideas behind this work.

This project was funded by the European Research Council under ERC Synergy Grant ECOGAL (grant 855130), led by Patrick Hennebelle, Ralf Klessen, Sergio Molinari and Leonardo Testi.
This work was granted access to HPC resources of CINES and
CCRT under the allocation x2020047023 made by GENCI (Grand
Equipement National de Calcul Intensif) and the special allocation 2021SA10spe00010. 
NB and RSK acknowledge computing resources provided by the Ministry of Science, Research and the Arts (MWK) of the State of Baden-W\"{u}rttemberg through bwHPC and the German Science Foundation (DFG) through grants INST 35/1134-1 FUGG and 35/1597-1 FUGG, and for data storage at SDS@hd funded through grants INST 35/1314-1 FUGG and INST 35/1503-1 FUGG.

\end{acknowledgements}

\section*{Data availability}

The data underlying this article are available in the Galactica Database at \url{http://www.galactica-simulations.eu}, and can be accessed with the unique identifier \textsc{\href{http://www.galactica-simulations.eu/db/STAR_FORM/TURBOX}{TURBOX}}.
Additional data and the source code used to run simulations and perform analysis will be shared on reasonable request to the corresponding author.
The code to compute the TS model and to do the plots of this paper is available at \url{https://gitlab.com/turbulent-support/ts_sfr_model}.

\bibliographystyle{aa}
\bibliography{global,lars}

\appendix

\section{Convergence with resolution}
\label{sec:conv}
\begin{figure}
    \centering
    \includegraphics[width=\linewidth]{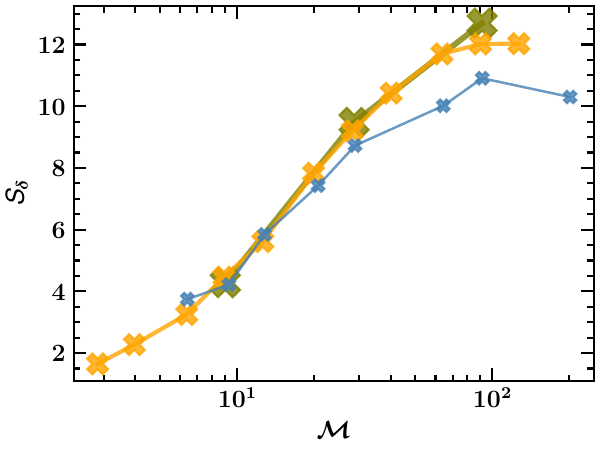}
    \includegraphics[width=\linewidth]{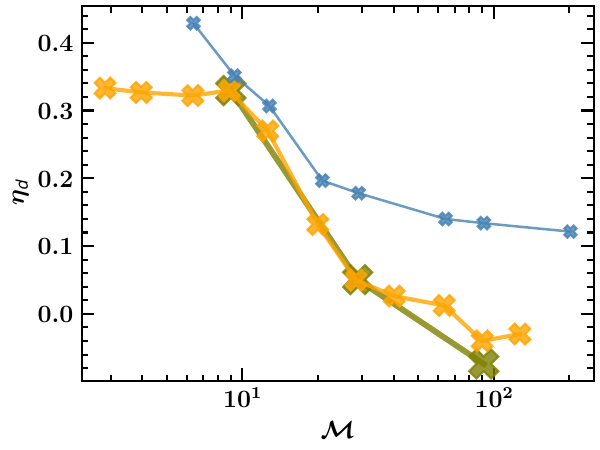}
    \includegraphics[width=\linewidth]{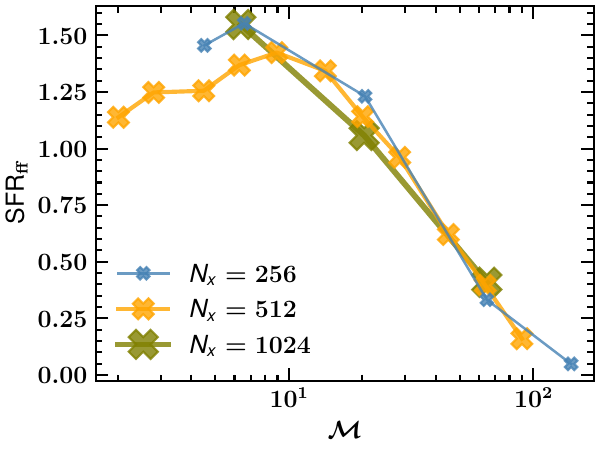}

    \caption{ Variance of the natural logarithm of the normalized density $S_\delta$ (top), $\eta_d$ parameter (middle) and SFR (bottom) as a function of Mach number, for resolution of 256$^3$ (blue, small markers), 512$^3$ (orange, medium-sized markers) and 1024$^3$ (green, big markers) for a driving compressibility $\chi = 0.5$.}
    \label{fig:conv}
\end{figure}

Most of the simulations were run at the resolution of (512$^3$), allowing for a very broad parameter study. 
In this section we look at more and less resolved simulations and see how the results are changed.
Our result mainly relies on the density PDF, the slope of the power spectrum of ln-density and the SFR.
Fig. \ref{fig:conv}, shows that at our fiducial resolution of $N_x = 512$, the variance of the natural logarithm of the normalized density $S_\delta$, the $\eta_d$ parameter and the SFR are converged, with no significant variation with the higher resolution of $N_x = 1024$. 
Slight deviations on $S_\delta$ and $\eta_d$ can be observed at very high Mach number ($\mathcal{M} \geq 100$). 
For $S_\delta$, no flattening of the curve is observed for the highest resolution, and it's quite clear the Mach number for which the curve start to flatten is smaller for lower resolution.
Such flattenings have been observed in many other simulations, as reported by \cite{hopkinsModelNonlognormalDensity2013} in their Figure 6. These are probably pure numerical artifacts.

\label{subsec:resolution_pspec}

\section{Power spectrum and scale-evolution of the PDF}
\label{sec:scale_evo}

\begin{figure*}
    \centering
    \includegraphics[width=\textwidth]{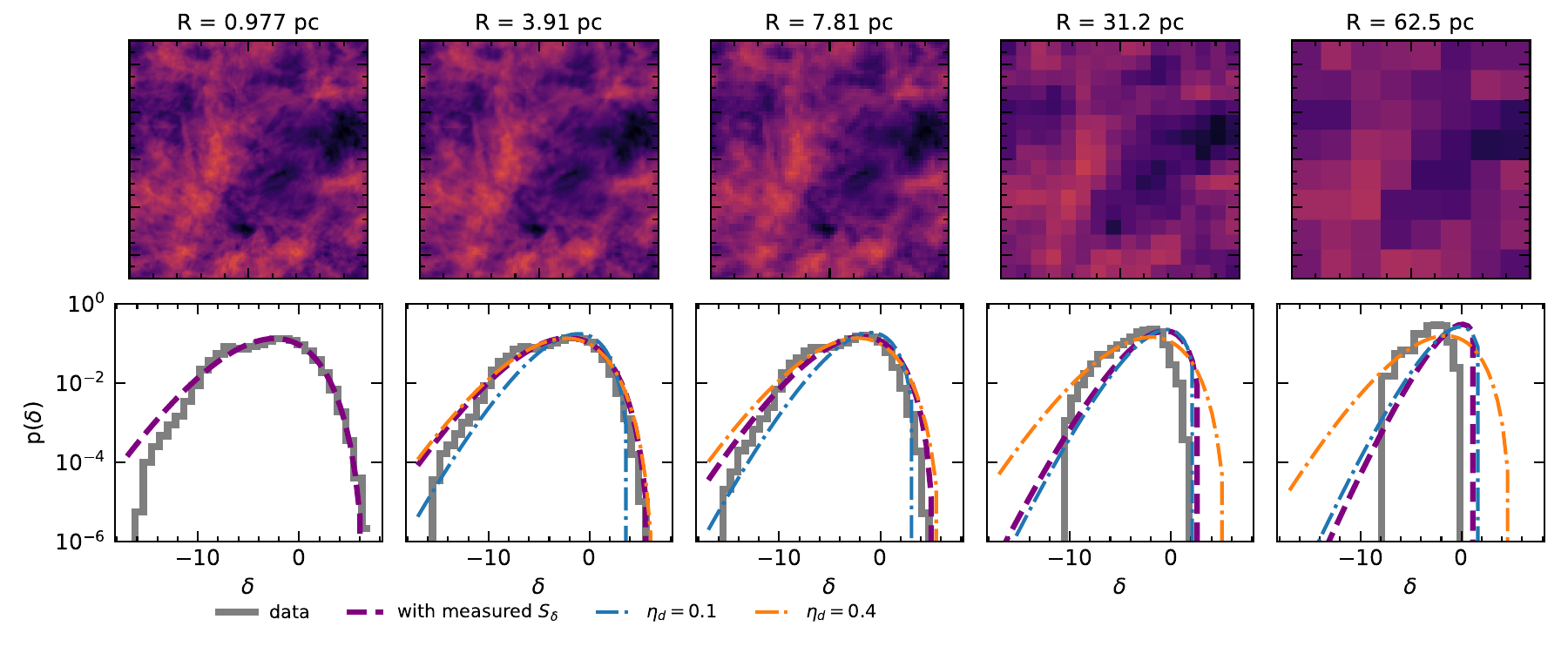}
    \caption{Top: Column density of the simulation L1000\_N1024\_comp0.5\_frms1e3 with the density averaged over a region of a given size $R$. The leftmost panel correspond to the resolution of the simulation. Bottom: PDF of $\delta$ when with the density being over a region of a given size $R$ (solid grey line). The dashed purple line is the Castaing-Hopkins PDF for the measured variance at this scale, while the blue and orange dashed-dot lines are the predicted values for the PDF using Eq.~\eqref{eq:S_R} for $\eta_d = 0.1$ and $\eta_d = 0.5$ respectively. }
    \label{fig:pdf_scale}
\end{figure*}

We use the value we computed for the slope of the log-density power spectra to get an estimate of the width of the log-density PDF at a given scale, using the formula introduced in \citetalias{hennebelleInefficientStarFormation2024}:
\begin{eqnarray}
S_\delta(R) &=& S_\delta \left( 1 - \left( \dfrac{R}{L_i} \right)^{2 \eta_d} \right).
\label{eq:S_R}
\end{eqnarray}
The variance $S_\delta(R)$ is the variance of the natural logarithm of the normalized density $\delta$ averaged over regions of size $R$, and holds crucial information to estimate the SFR in analytical models like the one presented in \citetalias{hennebelleInefficientStarFormation2024}. 
Using the fitted value of $\eta_d$ of Table \ref{tbl:turbox_simu} and Fig. \ref{fig:turbox_eta} within Eq.~\eqref{eq:S_R} would be very convenient. 
However, it can be noted in Fig.~\ref{fig:turbox_pspec} that the fit from which $\eta_d$ was computed do not cover the full range of scales in the simulation, and the spectrum at small scales is always steeper.
This has to be taken into account when using Eq.~\eqref{eq:S_R}, as shown by Fig. \ref{fig:pdf_scale}. Indeed, we see in that figure that one value of $\eta_d$ cannot reproduce the measured value of $S(R)$ measured at different scales in the simulation.

\section{Others metrics}

\begin{figure*}[ht]
    \centering
    \includegraphics[width=\textwidth]{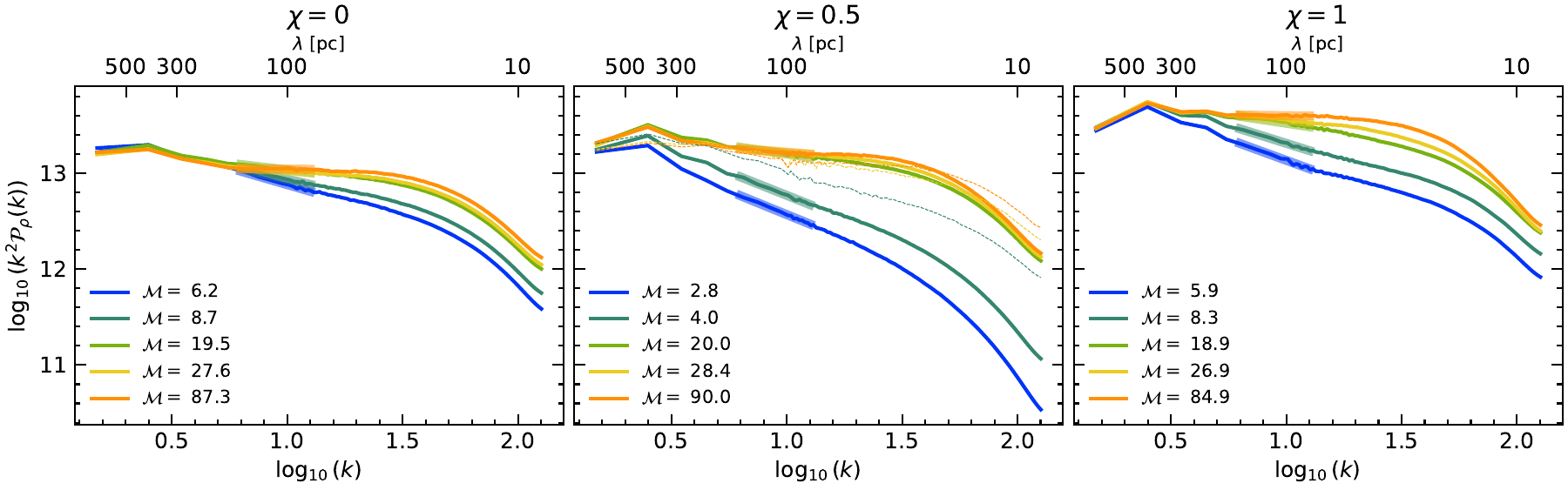}
    \includegraphics[width=\textwidth]{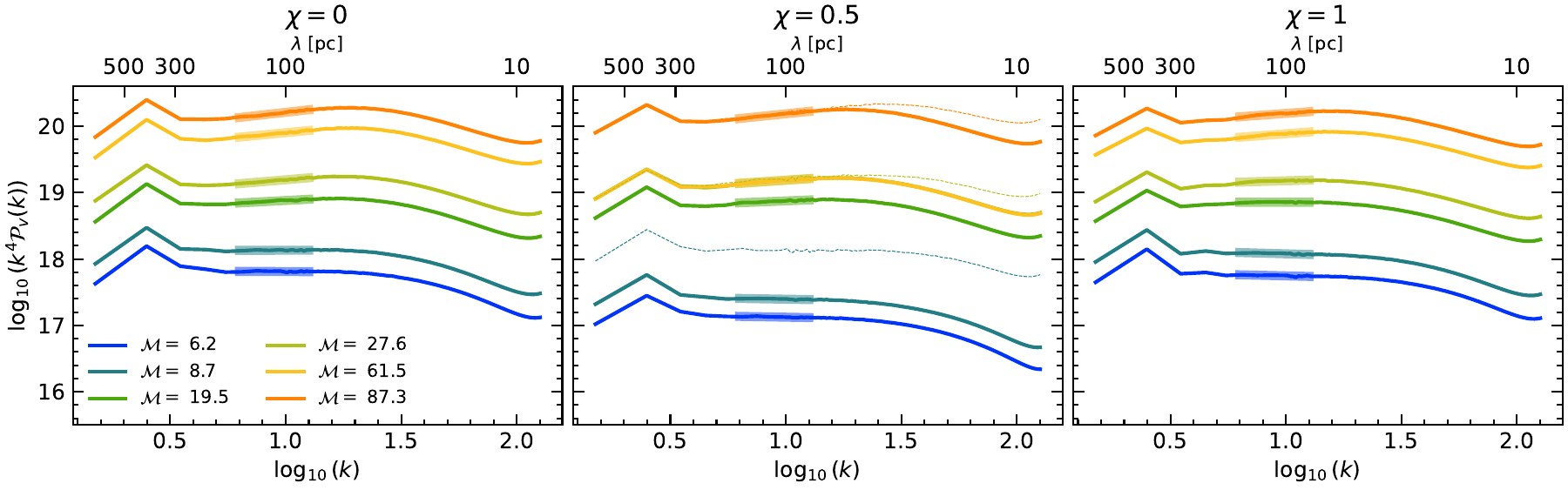}

    \caption{Power spectrum of the density (top) and the norm of the velocity (bottom) for the group L1000\_512 (solid lines) and  L1000\_1024 (dotted lines in the middle panel).}
    \label{fig:spect_rho_v}
\end{figure*}

For completeness, Fig. \ref{fig:spect_rho_v} features the power spectrum of density and velocity in our simulations.

\section{Smaller box size}

\begin{figure}
    \centering
    \includegraphics[width=\linewidth]{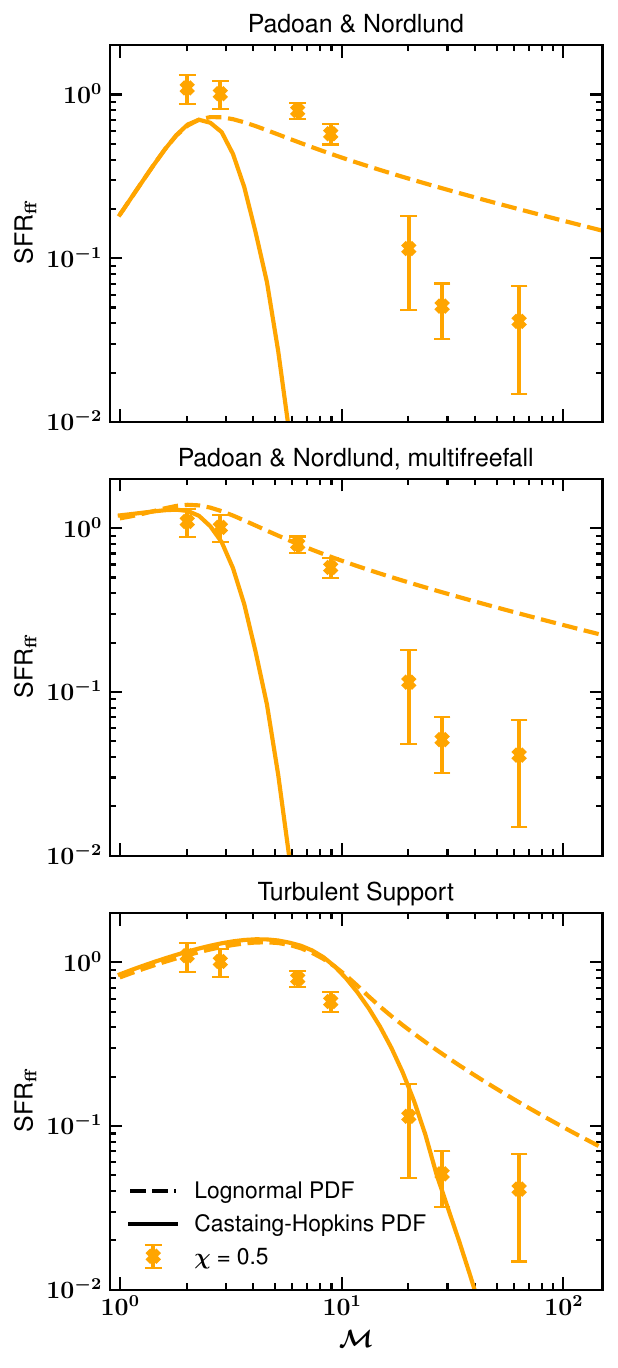}
    \caption{Same as Figure~\ref{fig:comparison_with_models} but with a box size of 200 pc and only for a driving compressibilty $\chi = 0.5$.}
    \label{fig:sfr_L200}
\end{figure}

We also ran simulations for a smaller box size with $L_\mathrm{box} = 200$ pc, close the typical galactic disk thickness (twice the scale height). Figure~\ref{fig:sfr_L200} shows the same comparison
as in Figure~\ref{fig:comparison_with_models}, with a driving compressibility $\chi = 0.5$.
The TS model used $L_i = 100$ pc, $\eta_d = 0.4$ and $y_\mathrm{cut} = 0.5$, and as with the larger box, is the only to feature a behaviour matching the simulations.
As predicted by the TS model, the transition between efficient and inefficient star formation regime occurs for a lower Mach number.

\end{document}